\documentclass[12pt]{article}

\usepackage[pdftex]{graphicx}
\usepackage[absolute]{textpos}

\newcommand{\dpar}      [2] {\frac{\partial #1}{\partial #2}}

\newcommand{\ndpar}     [3] {\frac{\partial^{#1} #2}{\partial #3^{#1}}}
\newcommand{\schr}{Schr\"o\-din\-ger}

\newcommand{\ivp}{initial-value problem}
\newcommand{\ist}{inverse scattering transform}
\newcommand{\ibvp}{initial-boundary value problem}
\newcommand{\D}{Dirichlet}
\newcommand{\N}{Neumann}
\newcommand{\R}{Robin}
\newcommand{\bc}{boundary condition}
\renewcommand{\b}{boundary}
\newcommand{\bs}{boundaries}
\renewcommand{\P}{periodic}
\newcommand{\eNss}{exact $N$-soliton solution}

\newcommand{\s}{soliton}
\newcommand{\sw}{solitary wave}
\newcommand{\phs}{phase shift}
\newcommand{\text}[1]{\mbox{#1}}
\newcommand{\sech}{\text {\,sech}}
\newcommand{\sgn}{\text {\,sgn}}
\newcommand{\eqref}[1]{(\ref{#1})}

\newcommand{\ds}{\displaystyle}
\renewcommand{\ss}{\scriptstyle}
\newcommand{\Z}{\bf {Z}}
\newcommand{\OL}{{\sf O}}

\title{The Nonlinear Schr\"{o}dinger Equation in the Finite Line}

\author{J. I. Ramos and F. R. Villatoro \\
    Departamento de Lenguajes y Ciencias de la Computaci\'on \\
    E. T. S. Ingenieros Industriales \\
    Universidad de M\'alaga \\
    Plaza El Ejido, s/n \\
    29013-M\'alaga  \\
    SPAIN
    }

\date{}
\begin{document}

\maketitle

\begin{abstract}

A numerical study of the nonlinear Schr\"{o}dinger (NLS) equation subject to
 homogeneous \D, \N\ and \R\ \bc s in the finite line is presented. The results
are compared with both the exact analytical ones for the \ivp\ (IVP) of the NLS
equation  and  the numerical ones for \P\ boundary conditions.  It is shown
that  initial solutions obtained by truncating the  \eNss\  of  the  IVP of the
NLS equation into a finite interval develop \sw s that behave as \s s, even
after collisions with the \bs. For periodic and  homogeneous \D\ and \N\ \bc s,
it is observed that the interaction between \s s and \bs\ is equivalent to the
collision between \s s in IVP or quarterplane problems. It is shown that for
homogeneous Robin boundary conditions, boundary layers that trap and delay the
soliton are formed at the boundaries. Phase diagrams for the \s\ amplitude at the
boundary points and for the \s's maximum amplitude show a recurrent phenomenon,
and are similar to those of the cubic Duffing equation. It is also shown that
the phase diagrams are strong functions of the parameter that defines the Robin
boundary conditions. A method of images, similar to the one  used in potential theory, is
developed for the NLS equation in the quarterplane with homogeneous \D\ and \N\
 \bc s at the finite boundary.

\vspace{1cm}

\end{abstract}

\section{Introduction}

\label{se:introduction}

The NLS equation has been used as a  model for the propagation of the envelope of
a wave packet in a weakly  nonlinear, dispersive  medium, and has found many
applications in physics, e.g., the propagation  of deep water gravity waves in
fluid mechanics, the propagation of Langmuir waves in plasma  physics and  the
self-focusing and self-modulation of trains of monochromatic waves in nonlinear
optics~\cite{Scot73,Ablo81,Roge82}. Its wide
applicability has an analytical justification since the NLS equation emerges
from almost all evolution equations characterized by a linear part which is
``dispersive" and a nonlinear part wich is ``analytical" when the interest lies
in the modulation of a carrier  wave due to weakly nonlinear
effects~\cite{Calo91}. Furthermore, the NLS equation is amongst the few nonlinear
partial differential equations that have  an IVP integrable by the
\ist~\cite{Ablo81,Zakh72}.

Most of the numerical studies on the NLS equation have  been concerned with the
IVP or infinite line problem which was truncated to a finite one and subjected
to periodic~\cite{Weid86,Sanz84,Taha84}, homogeneous Dirichlet~\cite{Akri93}
 or homogeneous Neumann~\cite{Grif84,Argy87,Sham90} \bc s.  However, experiments
have mainly dealt with the generation of solitons, i.e., boundary-generated
solitons, which must be  modelled by  an \ibvp\ in the quarterplane.   These
\ibvp s for nonlinear integrable systems in the quarterplane are also called
`forced integrable  systems'~\cite{Kaup85} because the \bc s can be viewed as
forces that act at  the boundaries.

Despite the great interest that the NLS equation has received in the past two
decades through  analytical and numerical studies, its \ibvp\ in semi-infinite
and finite lines is still a subject of current research because of   serious
analytical difficulties~\cite{Ablo91a}.  Three different approaches have been
used to study the  semi-infinite line, \ibvp\  of nonlinear, integrable
equations, here referred to as the quarterplane problem, and to analyze the
propagation of boundary-generated  solitons~\cite{Chu90}. The first approach is
based on the use of numerical methods to study the soliton's generation and
formation in quarterplane problems subject to nonhomogeneous boundary conditions
at the finite boundary. For example, Chu~\cite{Chu83} and Chou and
Chu~\cite{Chou90} studied boundary-generated solitons for the Korteweg-de~Vries
and  Boussinesq equations, respectively.  The second approach is based on the
\ist\ and employs ad hoc asumptions in order to model known numerical results.
Kaup~\cite{Kaup85} and Kaup and Hansen~\cite{Kaup86} used this second approach
to study the quarterplane problem of the NLS equation. The third approach
is based on  extensions of the \ist\ and employs  nonlinear sine-Fourier
transforms. This approach has been applied to the NLS equation by Ablowitz and
Segur~\cite{Ablo75} who considered homogeneous Dirichlet and Neumann \bc s at
the finite boundary, and has been generalized by Fokas~\cite{Foka89a} and Bikbaev
and  Tarasov~\cite{Bikb91} for homogeneous Robin \bc s. Bikbaev and
Tarasov~\cite{Bikb91} used B\"acklund transformation methods in their
analysis, while Fokas~\cite{Foka89a} has  shown that the scattering data for
the  nonhomogeneous Robin \bc\ problem is governed by a nonlinear,
integro-differential equation, although he was unable to obtain its solution.
Fokas and Ablowitz~\cite{Foka89b} have also shown that the infinite-line, forced
NLS equation may be used to study the same equation in some quarterplane
problems by properly choosing the forcing term.

The \ibvp\ of the NLS equation in the finite line has received
much less attention than  IVP and quarterplane problems.  In fact, only the
periodic problem seems to have  been  studied in detail.  Ma and
Ablowitz~\cite{Ma81} have obtained analytically  finite N-band potential
solutions of the NLS equation subject to periodic boundary conditions by means
of the  periodic \ist.  Osborne~\cite{Osbo93} has developed a  fast numerical
version of the \ist, while Boyd~\cite{Boyd90} has studied imbricate series and
 polycnoidal waves of the periodic NLS equation. The non-periodic,
finite-line, \ibvp\
 for other integrable, nonlinear evolution equations has also received
little attention. Christiansen~\cite{Chri85} and Sirovich et
al.~\cite{Siro90} have studied numerically the \N\ problem for the
sine-Gordon and the Ginzburg-Landau equations, respectively.

In this paper,  a numerical  study of the \ibvp\ of the NLS equation in a
finite interval subject to homogeneous \D, \N\ and \R\ \bc s at both boundaries
is presented. The results are compared with both the exact analytical ones for
the IVP of the NLS equation~\cite{Zakh72} and with the numerical results of the
NLS equation subject to periodic boundary conditions. Particular attention is
paid to the initial conditions and their mathematical compatibility with the
boundary ones, and to the propagation of  \s s and their interactions with  the
\bs\ which are compared with those that result from the mutual interactions
between \s s in the infinite spatial line, i.e., with those corresponding to
the IVP of the NLS equation. It must be noted that collisions between \s s and
\bs\ have seldom been considered previously, even for other integrable, nonlinear
equations except for Christiansen~\cite{Chri85} who studied  the \N\
boundary-value problem for the sine-Gordon equation in Josephson tunnel
junctions, Mirie and Su~\cite{Miri82} who analyzed numerically the \N\
boundary-value problem for the Korteweg-de Vries equation, and  Chou and
Chu~\cite{Chou90} who considered   the
reflection of solitons from a wall for the Boussinesq equation.

The paper has been organized as follows. The one-dimensional NLS equation and
its analytical solution for the IVP is presented in Section~\ref{se:formulation}
together with the homogeneous boundary conditions at the ends of the finite line
used throughout this paper. The second-order, accurate, finite difference method
employed to solve the NLS equation is described in detail in
Section~\ref{se:numerical}. Section~\ref{se:quarterplane} is devoted to
quarterplane plane problems subject to homogeneous \D\ and \N\ at the finite
boundary. In that section, it is  shown
that the collision between a \s\ and a homogeneous \N\ \bc\ is equivalent to the
collision  between two identical \s s  of opposite velocities, whereas the
collision  of a soliton with a homogeneous \D\ \bc\ is equivalent to the
collision between two  \s s of  opposite amplitudes and velocities. In
Section~\ref{se:quarterplane}, $N$-soliton solutions are obtained for the \D\
and \N\ quarterplane problems. Section~\ref{se:finiteline} is entirely devoted
to the study of the NLS equation in the finite line subject to homogeneous \D,
\N, \R\ and \P\ \bc s  which are mathematically compatible
with the initial conditions. In that section, it is shown that the
initial conditions evolve into \sw s which have a \s -like shape  and  retain
their shape after collisions with the boundaries except for a  \phs, i.e.,
they behave as \s s. In Section~\ref{se:finiteline}, it is also shown that the
collisions of the solitons with the boundaries exhibit a recurrent phenomena
which is analyzed  by means of phase  diagrams for the \s\ amplitude at the
boundary points  and for the \s\ maximum amplitude. These diagrams show large
topological changes as  the parameter that  controls the \R\ \bc s is varied.
They also show that the soliton amplitude in both IVP and   periodic problems
behaves as a nonlinear cubic Duffing equation which may exhibit  chaotic
behaviour under certain forcings~\cite{Ablo91b}.

\section{Formulation of the Problem}

\label{se:formulation}

The one-dimensional NLS equation in dimensionless form can be written as
    \begin{equation} \label{eq:nse} \label{eq:nlse}
iu_t+u_{xx}+q |u|^2 u=0, \qquad x \in {\cal D} \text{\ and \ } t \geq 0
    \end{equation}
where $t$ is time, $x$ is the spatial Cartesian coordinate,  $u$ is the complex
amplitude, $q$ is a real number, $i= \sqrt{-1}$, the subscripts denote
partial differentiation, and $\cal D$ denotes the domain of definition of the
equation, i.e.,  the whole real line for the IVP, a semi-infinite line
with one boundary  located at $x$=0, i.e., ${\cal D} \equiv [0,\infty)$ for
quarterplane problems, or a finite interval, ${\cal D} \equiv [-L,L]$, with two
\bc s  for two-point, \ibvp s. Note that, without loss of generality,  a
symmetric interval has been chosen since the NLS equation is invariant under
both translations and mirror reflections in $x$.

The NLS equation in the infinite line has analytical solutions which can be
determined by means of the inverse scattering transform. For example, the
exact $N$-soliton solution of the NLS equation in the infinite line  was
obtained  by Zakharov  and Shabat~\cite{Zakh72}. Gordon~\cite{Gord83}
essentially  reduced by  half the number of equations obtained by Zakharov and
Shabat and  redefined their parameters to give them clearer meanings.

The exact $N$-soliton solution of the NLS equation  given by
Gordon~\cite{Gord83} can be written as
    \begin{equation} \label{eq:Nss:sum}
u(x,t) = \frac {1}{ \sqrt {q} } \sum_{j=1}^N u_j(x,t),
    \end{equation}
where $u_j$ is the solution of the following  system of linear
 algebraic equations
    \begin{equation} \label{eq:Nss:syst}
\sum_{k=1}^N M_{jk} u_k = \sum_{k=1}^N \frac{\gamma_j^{-1} +
\gamma_k^*}{\lambda_j + \lambda_k^*} u_k = 1, \qquad j=1, 2, \ldots N,
    \end{equation}
and
    \begin{equation} \label{eq:Nss:deflam}
\lambda_j=A_j+iV_j / \sqrt{2},
    \end{equation}
    \begin{equation} \label{eq:Nss:defgam}
\gamma_j=\exp \left[ \lambda_j(x-x_{j0}) / \sqrt{2} + i \lambda_j^2 t/2 + i
                     \phi_{j0} \right],
    \end{equation}
where the four real parameters
$\{A_j, V_j, x_{j0}, \phi_{j0}\}$
 denote the amplitude, velocity, initial location of the maximum
amplitude and initial phase,  respectively, of the $j$-th soliton.

Equation~\eqref{eq:Nss:sum} can be expanded in the neigborhood of a soliton well
separated from the others to yield~\cite{Gord83}
    \begin{equation}
u_j(x,t)=\frac {A_{j}}{\sqrt {q}} \sech \{ \frac{A_{j}}{\sqrt{2}} ( x-x_{j}) +
\frac{q_j}{\sqrt{2}} \} \exp\{i[ \phi_{j} +\Phi_{j} ] \},
    \end{equation}
where
\begin{equation}
x_j = x_{j0} + V_{j}t , \qquad
\phi_j = \frac{V_{j}}{2}(x-x_{j0})+(A_{j}^2-\frac{V_{j}^2}{2})\frac{t}{2} +
\phi_{j0},
 \end{equation}
and the presence of the other distant solitons introduces a displacement,
$q_j/{\sqrt{2}}$, and a phase shift, $\Phi_j$, given by
    \begin{equation}
q_j + i \Phi_{j} = \sum_{k \neq j} \sgn\, \left( x_{k}-x_{j} \right)
  \log \left[ \frac {A_j+A_k + i {(V_j-V_k)}/{\sqrt{2}}}
                    {A_j-A_k + i {(V_j-V_k)}/{\sqrt{2}}} \right],
    \end{equation}
where the sign function, $\sgn\,(x)$, is equal to one, zero and minus one, for
positive, zero and negative $x$, respectively.

Notice that for the 1-soliton case,
Eqs.~\eqref{eq:Nss:sum}--\eqref{eq:Nss:defgam} yield
    \begin{equation} \label{eq:1ss}
u(x,t) = \frac{A}{\sqrt {q}} \sech \{ \frac{A}{\sqrt{2}} [ (x-x_0)- Vt] \}
  \exp \{ \frac{i}{2} [ V(x-x_0) + (A^2-\frac{V^2}{2})t ] + i \phi_0 \}.
    \end{equation}

It is well known that the IVP of the NLS  describes a completely integrable
Hamiltonian system with the  scattering data as action-angle variables and an
infinite set of invariants. The most important invariants of the IVP of the NLS
equation  are the first, second and third ones.  The first, known as wave mass
or `number of particles', is
    \begin{equation} \label{eq:inv1}
I_1 = \int_{-\infty }^\infty  {|u|^2\,dx}.
        \end{equation}
The second invariant represents the total momentum in the Hamiltonian
formalism and is given by
    \begin{equation} \label{eq:inv2}
I_2 = i \int_{-\infty }^\infty  {\left(uu_x^* - u^*u_x \right) \,dx},
        \end{equation}
where the asterisk  denotes complex conjugate.
The third invariant is the total energy or Hamiltonian, i.e.,
    \begin{equation} \label{eq:inv3}
I_3 = \int_{-\infty }^\infty {\left(\left|u_x\right|^2-\frac{q}{2}|u|^4\right)
\,dx}.
        \end{equation}

It may be easily shown from Eq.~\eqref{eq:nlse} that
\begin{equation} \label{eq:evolmass}
\dpar{m}{t} + \dpar{M}{x} = 0,
\end{equation}
where $m=u^*u=|u|^2$ and $M=i(uu_x^* - u^*u_x)$ denote the mass density and
linear momentum density, respectively.

In this paper, the analytical solution of the NLS equation in the
infinite line is compared with the numerical solution of the same
equation in finite lines subject to the following homogeneous,
 boundary conditions
    \begin{equation} \label{eq:nse:r}
u(x,t) + \gamma u_x(x,t) = 0, \qquad x \in {\partial\cal D}
                    \text{\ and \ } t \geq 0,
    \end{equation}
where $\gamma$=0 and $\infty$ correspond to homogeneous Dirichlet and Neumann
boundary conditions, respectively, whereas finite values of $\gamma \ne 0$
correspond to homogeneous Robin boundary conditions at both boundaries. In
addition, the NLS equation is also solved in a finite line subject to the
following periodic boundary conditions
    \begin{equation} \label{eq:nse:p}
\ndpar{n}{u}{x}(x,t) = \ndpar{n}{u}{x}(x+2kL,t),
  \qquad \forall n \geq 0, \quad k \in {\Z},
         \quad x \in {\cal D} {\equiv [-L,L] \text{\ \ and \ } t \geq 0}.
    \end{equation}
Hereon, the NLS equation subject to periodic and homogeneous Dirichlet, Neumann
and Robin boundary conditions will be referred to as the periodic, Dirichlet,
Neumann and Robin problems, respectively.

\section{Numerical Scheme Used in the Simulations}

\label{se:numerical}

For the numerical integration of Eq.~\eqref{eq:nlse} in the interval
$[-L,L]$, the Crank-Nicolson finite difference method has been used. This
method can be written as
    \begin{equation} \label{eq:cne}
i \frac{U_j^{n+1}-U_j^n}{\Delta t}
+ \frac{1}{\Delta x^2} \delta^2  U_{j}^{n+1/2}
+ q \left| U_{j}^{n+1/2} \right|^2 U_{j}^{n+1/2}
= 0,
    \end{equation}
where
\[ \delta ^2 U_j^n = U_{j+1}^n - 2 U_j^n + U_{j-1}^n , \qquad
   U_{j}^{n+1/2} = \frac{U_j^{n+1}+U_j^n}{2},
\]
and
 \[ U_j^n = u(-L+j \Delta x, n \Delta t),
\] \[ j=0, 1, \ldots N , \quad n \geq 1, \quad \Delta x = 2L/N,
      \quad \Delta t > 0,
\]
where  ($N$+1) is the number of grid points; $\Delta x$ is the spatial step
size; and, $\Delta t$ is the time step. This scheme is second-order accurate in
both space and time, i.e., \mbox { $\OL (\Delta t^2,\Delta x^2)$}.

The discrete boundary conditions are as follows  (cf. Eq.~\eqref{eq:nse:r}). For the \D\ \bc s, i.e.,
$\gamma = 0$,
    \begin{equation} \label{eq:nbc:d}
U_0^{n+1/2} = U_N^{n+1/2} = 0.
    \end{equation}
For the \N\ \bc s, i.e.,
$\gamma = \infty$,
    \begin{equation} \label{eq:nbc:n}
\frac {U_{1}^{n+1/2}-U_{-1}^{n+1/2}} {2 \Delta x}
= \frac {U_{N+1}^{n+1/2}-U_{N-1}^{n+1/2}} {2 \Delta x} = 0.
    \end{equation}
For the \R\ \bc s, i.e.,
$\gamma \ne 0$ and $\gamma \ne \infty$,
    \begin{equation} \label{eq:nbc:r}
U_0^{n+1/2} + \gamma \frac {U_{1}^{n+1/2}-U_{-1}^{n+1/2}} {2 \Delta x} =
U_N^{n+1/2} + \gamma \frac {U_{N+1}^{n+1/2}-U_{N-1}^{n+1/2}} {2 \Delta x} = 0.
    \end{equation}
For the \P\ \bc s (cf. Eq.~\eqref{eq:nse:p})
    \begin{equation} \label{eq:nbc:p}
U_{N+1}^{n+1/2} = U_1^{n+1/2} ,\qquad U_{-1}^{n+1/2} = U_{N-1}^{n+1/2}.
    \end{equation}
Note that two fictitious points have been introduced at $j$=$-1$ and
$j$=$N$+1,  except in the \D\ problem, in order to preserve the second-order,
spatial accuracy of the  Crank-Nicolson method.

It can  be easily shown that the Crank-Nicolson method is linearly stable and
nonlinearly stable by means of an energy method. The existence and  convergence
of the discrete solutions  of the discretized IVP of the NLS equation have been
proved  by  Ben-Yu~\cite{Ben84} and his results  can be  extended to a finite
interval with  homogeneous \bc s.

An iterative, Newton-Raphson method  has been used to solve the nonlinear
algebraic system of equations (cf. Eq.~\eqref{eq:cne}). The convergence of this
method for finite-line problems can be proven using  similar methods to those
employed by Akrivis~\cite{Akri93}. The diagonally dominant, tridiagonal system
of linear algebraic equations that result from the Newton-Raphson  method has
been solved by a 2$\times$2 block-oriented version of the Thomas algorithm. For
periodic problems,  a natural optimization  of the Gaussian elimination
technique for cyclic-tridiagonal systems has been used before  employing the
Thomas algorithm~\cite{Taha84}.

The Crank-Nicolson method is conservative since it preserves a discrete
equivalent of the mass (cf. Eq.~\eqref{eq:inv1}), i.e., the $L^2$ norm
of the  solution, for the discrete, infinite line IVP  and for  finite
line problems subject to \P\ or homogeneous \D\ \bc s.  For homogeneous \N\ and
\R\ \bc s, the mass is  nearly preserved when the solitons are  far away
from, i.e., they are not affected by, the  boundaries or when a small $\Delta t /
\Delta x^2$ is  employed in the numerical calculations. This can be shown as
follows. The imaginary part of the sum (in $j$)  of the product
of Eq.~\eqref{eq:cne} times
 $(U_j^{n+1} + U_j^n)^{\ds *}$ yields
    \begin{equation} \label{eq:ninv1:pd}
\| U^{n+1} \| = \| U^n  \| =\| U^0 \| , \qquad n \geq 1
    \end{equation}
for the discrete, infinite line problem and for both the \P\ and the
\D\ problems, and
    \begin{eqnarray} \label{eq:ninv1:nr}
&&\|U^{n+1}\| - \|U^n\| - \frac {\Delta t} {2 \Delta x^2}
{\cal I}m [ (U_1^{n+1} + U_1^n) (U_0^{n+1} + U_0^n)^{\ds *} + \nonumber \\
&& \phantom {\|U^{n+1}\| - \|U^n\| - \frac {\Delta t} {2 \Delta x^2}
{\cal I}m [}
 \mbox{} + (U_{N-1}^{n+1} + U_{N-1}^n) (U_N^{n+1} + U_N^n)^{\ds *}
   ] = 0
    \end{eqnarray}
for the \N\ and \R\ problems, where ${\cal I}m$ denotes  imaginary part and
$\| U_j^n \|$ represents the $L^2$ norm of $U_j^n$.

In previous numerical papers, e.g.,~\cite{Weid86}--\cite{Sham90}, only the mass
(cf. Eq.~\eqref{eq:inv1}) and energy (cf. Eq.~\eqref{eq:inv3}) have been studied in
order to assess the conservation properties of the numerical methods used to
solve the NLS equation in infinite lines which were truncated to finite ones
subject to either homogeneous Dirichlet, homogeneous Neumann or periodic
boundary conditions.  In the finite line problems considered in this paper, the
mass, total momentum and total energy of the NLS equation may not be conserved,
i.e., they may not be invariants of the NLS equation in the finite line.
Furthermore, the total momentum (cf. Eq.~\eqref{eq:inv2}) is also calculated
here since it can be used to  assess the influence of the discretization on the
velocity of the soliton which is equal to the ratio of the total momentum to the
total mass for a soliton in the IVP of the NLS equation. Moreover, it is a good
numerical practice to calculate the total momentum in order to check the effect
of the spatial  discretization on the evaluation of both the first-order spatial
derivative and  the total energy (cf. Eq.~\eqref{eq:inv3}). The
 total momentum and the total energy of the NLS equation are nearly preserved by
the Crank-Nicolson  scheme (cf. Eq.~\eqref{eq:cne}) for the IVP of the NLS
equation. In  this paper, the mass, total momentum and total energy are evaluated
numerically  by means of  the  Simpson's rule which is fourth-order
accurate~\cite{Rals78}.

In the numerical simulations  of finite line problems presented in this paper,
the initial condition corresponding to the \eNss\   truncated to a finite
interval has  been used. However, this initial condition  (cf.
Eq.~\eqref{eq:1ss}) is mathematically
incompatible with the boundary conditions (cf. Eq.~\eqref{eq:nse:r}). In order to
avoid mathematical incompatibilities, the following initial condition was used
in the simulations presented in this paper
    \begin{equation} \label{eq:u0:form}
U_0(x) = u(x,0)+ax^2+bx+c,
    \end{equation}
where the values of $a$, $b$ and $c$ were determined in such a manner that the
initial condition, i.e., $U_0(x)$, satisfied the homogeneous boundary
conditions, and $u(x,0)$ is given by Eq.~\eqref{eq:1ss}. It may be easily shown
that, for the Dirichlet problem,
    \begin{equation} \label{eq:u0:dir}
a=0, \qquad
b=\frac{i}{L}{\cal I}m[u(-L,0)], \qquad
c=-{\cal R}e[u(-L,0)],
    \end{equation}
for the Neumann problem,
    \begin{equation} \label{eq:u0:neu}
a=\frac{1}{2 L}{\cal R}e[u_x(-L,0)], \qquad
b=-i \, {\cal I}m[u_x(-L,0)], \qquad
c=0,
    \end{equation}
and, for the Robin problem,
    \begin{eqnarray} \label{eq:u0:rob}
a&\!=\!&0,\qquad
  b=\frac{1}{L}
    \left( \gamma {\cal R}e[u_x(-L,0)] + i \, {\cal I}m[u(-L,0)]
    \right), \nonumber \\
\lefteqn { c = \frac{1}{L}
  \left\{ -(1+i)\gamma^2 {\cal R}e[u_x(-L,0)] +
           (1-i) \gamma {\cal I}m[u(-L,0)]
  \right\} - }\nonumber \\
& & \mbox{}-{\cal R}e[u(-L,0)] - i \gamma {\cal I}m[u_x(-L,0)],
    \end{eqnarray}
where ${\cal R}e$ denotes  real part. It must be noted that the
correction to the exact solution of the IVP of the NLS equation used in this
paper to avoid mathematical incompatibilities between the initial and boundary
conditions in finite line problems is not unique, and that other corrections
different from the parabolic one employed here may be used. However, the
difference between  the parabola chosen here and other possible functions is
a minor one because the  corrections introduced by the translation of the
initial conditions are on the order of $\exp(-A L)$ and, for  say, $A$=1 and
$L\geq  20$, these differences are very small compared with the soliton's
amplitude. Furthermore, simulations  have shown that the solutions of the NLS
equation with and  without  translation of the initial conditions  nearly yield
the same results except for the  small numerical errors introduced by the
discretization of the NLS equation and boundary conditions.

Symmetric, second-order accurate in space, computational
molecules were employed  at all the spatial grid points in
the periodic problem, while  asymmetric, three-point, second-order accurate
stencils were introduced at the boundary  points for the Dirichlet, Neumann
and Robin problems in order to evaluate the total momentum and the total energy.

\section {The NLS Equation in the Quarterplane}

\label{se:quarterplane}

In this section, the NLS in semi-infinite lines subject to homogeneous
Dirichlet or Neumann boundary conditions at $x$=0 is considered in order to
analyze the interaction of the soliton with the finite boundary and explain some
of the numerical results which will be presented in Section~\ref{se:finiteline}
regarding the interaction of solitons with the boundaries in finite line
problems.

\subsection {The \D\ Problem in the Quarterplane}

\label{se:Dimages}

In this section, it is shown that the interaction  between a soliton with
parameters $\{A_j, V_j, x_{j0},\phi_{j0}\}$ in the semi-infinite line and a
homogeneous Dirichlet boundary condition at $x=0$  is identical to the
interaction between that soliton and another soliton, i.e., its image with
respect to the \D\ boundary, of
parameters $\{A_j, -V_j, -x_{j0}, \pi + \phi_{j0}\}$.  It must be noted that
the  soliton image may also be represented by the  parameters $\{-A_j, -V_j,
-x_{j0}, \phi_{j0}\}$ which are  the ones employed in this paper. For the sake
of convenience in what follows, $u_s\equiv\{A, V,
x_{0},\phi_{0}\}$ and
 $\tilde{u}_s\equiv\{-A, -V,$ $-x_{0},$ $\phi_{0}\}$ will be used to define the
soliton in the quarterplane and its image with respect to a Dirichlet
boundary. In the case of several solitons in the quarterplane, the
subscript $j$ will be used to denote the $j$-th soliton.

Equations~\eqref{eq:Nss:deflam} and~\eqref{eq:Nss:defgam} yield the following
result at the boundary, i.e., at $x$=0,
    \begin{equation} 
\lambda=A+i V /\sqrt{2} =-\tilde{\lambda}, \qquad
\gamma=\exp ( -\lambda x_0 /\sqrt{2} + i \lambda^2 t /2 + i \phi_0 )
      = \tilde{\gamma},
    \end{equation}
while Eq.~\eqref{eq:Nss:syst} yields
    \begin{equation} 
\left( {\gamma ^{-1}+\gamma ^*} \right)
 \left( \begin {array}{cc}
   \ds \frac{1}{\lambda + \lambda ^*} & \ds \frac{1}{\lambda -\lambda ^*}\\ & \\
   \ds \frac{1}{-\lambda +\lambda ^*} & \ds \frac{1}{-\lambda -\lambda ^*}
 \end{array} \right)
 \left( \begin {array}{c} u_s \\ \tilde{u}_s \end{array} \right)
 = \left( \begin {array}{c} 1 \\ 1 \end{array} \right),
    \end{equation}
whose solution is the trivial one, i.e.,
    \begin{equation} 
 u (0,t) = u_s + \tilde{u}_s = 0.
    \end{equation}
Therefore, the superposition of a soliton which is the solution of the IVP of
the NLS equation and its image with respect to a Dirichlet boundary is a
solution of the NLS equation in the quarterplane with homogeneous Dirichlet
boundary conditions at $x$=0.

In order to show that the method of images presented
in previous paragraphs for two solitons is also valid for $N$ solitons in the
quarterplane with homogeneous Dirichlet boundary conditions at $x$=0, consider
the \eNss\ given by  Gordon~\cite{Gord83} for the IVP of the NLS equation (cf.
Eqs.~\eqref{eq:Nss:sum}--\eqref{eq:Nss:defgam}) and  apply the following
induction principle. Assume that  the  superposition of
$(N$--$1)$-\s s $\{u_{sj}\}$ and their images $\{\tilde{u}_{sj}\}$ is zero at
the \b\ in a process of soliton-image pair cancellation. In order to show that
the addition of another soliton and its image to the $(N$--$1)$-\s s and their
images is also zero, we first introduce the following notation
    \begin{equation} 
\eta_{jk}  =  \gamma_j^{-1} + \gamma_k^* , \qquad
\mu_{jk}^{-1}  =  \lambda_j + \lambda_k^* , \qquad
\nu_{jk}^{-1}  =  \lambda_j - \lambda_k^*,
    \end{equation}
and note that, for the homogeneous \D\ \bc\ at $x$=0,
    \begin{equation} 
\lambda_j = - \tilde \lambda_j , \qquad \gamma_j = \tilde \gamma_j.
    \end{equation}
From the induction hypothesis,
    \begin{equation} 
\tilde u_{sj} = - u_{sj}, \qquad {\rm for\ } j=1 , 2 , \ldots N-1,
    \end{equation}
while, from Eq.~\eqref{eq:Nss:syst},
the following linear  system of  $2 N$-equations must be satisfied
    \begin{eqnarray} 
\sum_{k=1}^{N-1} \eta_{jk} (\mu_{jk}-\nu_{jk}) u_{sj}
  + \eta_{jN} (\mu_{jN} u_{sN} + \nu_{jN} \tilde{u}_{sN} ) &=& 1, \\
\sum_{k=1}^{N-1} \eta_{jk} (-\nu_{jk}+\mu_{jk}) u_{sj}
  + \eta_{jN} (-\nu_{jN} u_{sN} - \mu_{jN} \tilde{u}_{sN} ) &=& 1,  \\
\sum_{k=1}^{N-1} \eta_{Nk} (\mu_{Nk}-\nu_{Nk}) u_{sN}
  + \eta_{NN} (\mu_{NN} u_{sN} + \nu_{NN} \tilde{u}_{sN} ) &=& 1, \\
\sum_{k=1}^{N-1} \eta_{Nk} (-\nu_{Nk}+\mu_{Nk}) u_{sN}
  + \eta_{NN} (-\nu_{NN} u_{sN} - \mu_{NN} \tilde{u}_{sN} ) &=& 1,
    \end{eqnarray}
for $j=1,2,\ldots,N-1$.

Substraction of the last two equations yields
    \begin{equation} 
(\mu_{NN}+\nu_{NN})(u_{sN} + \tilde{u}_{sN}) = 0,
    \end{equation}
which indicates that the  introduction of a new soliton is compensated by its
image. Therefore, the set of $N$ solitons and of their $N$ images satisfies
the homogeneous Dirichlet  \bc\ at $x$=0.

In order to illustrate the method of images developed in this section, the
interaction of two  solitons of $A_1$=$V_1$=$-A_2$=$-V_2$=$q$=1, $x_{01}$=0,
$x_{02}$=100, and $\phi_{01}$=$\phi_{02}$=0 is illustrated in
Figure~\ref{fg:2ss:dirichlet}. The results presented in
Figure~\ref{fg:2ss:dirichlet} simulate the interaction of a soliton with a
homogeneous Dirichlet boundary condition at $x$=50 in a quarterplane problem,
and indicate that the amplitude of both solitons increases as they approach
$x$=50 and that both solitons rebound from each other with the same amplitude
and speed as those prior to their collision or interaction with the finite
boundary.

\subsection {The \N\ Problem in the Quarterplane}

\label{se:Nimages}

In this section, we show that the interaction of a soliton with a Neumann
boundary in a quarterplane problem is  identical to the collision between two
solitons of the same amplitude and  initial phase, but of opposite velocities.
For a  soliton with parameters $\{A_j, V_j, x_{j0}, \phi_{j0}\}$,  its image with
respect to  a \N\ boundary located at $x=0$ has the parameters  $\{A_j, -V_j,
-x_{j0}, \phi_{j0}\}$. Without loss of generality, consider a  semi-infinite
line problem with the homogeneous \N\ \bc\  at $x=0$, and  a soliton with
parameters  $u_s\equiv\{A, V, x_{0}, \phi_{0}\}$ and its \N-image
$\tilde{u}_s\equiv\{A, -V, -x_{0}, \phi_{0}\}$.

From Eqs.~\eqref{eq:Nss:deflam} and~\eqref{eq:Nss:defgam},
    \begin{equation} 
\lambda=A+i V /\sqrt{2}=\tilde{\lambda}^*,
    \end{equation}
    \begin{eqnarray} 
\gamma & = & \exp ( \lambda (x-x_0) /\sqrt{2} + i\lambda^2 t/2 + i\phi_0 ) =
  e^{\lambda x /\sqrt{2}} \zeta, \\
\tilde{\gamma} & = & \exp ( \lambda^*(x+x_0) /\sqrt{2} + i\lambda^{*2} t/2 +
i\phi_0 ) =
  e^{\lambda^{\ss *} x /\sqrt{2}} \zeta^{*-1},
    \end{eqnarray}
and Eq.~\eqref{eq:Nss:syst} becomes
    \begin{equation} 
M(x,t) =
 \left( \begin {array}{cc}
  \ds\frac { e^{-\lambda x /\sqrt{2}} \zeta^{-1} +
                 e^{\lambda^{\ss *} x /\sqrt{2}} \zeta^{*} }
          { \lambda +\lambda ^* } &
  \ds\frac { e^{-\lambda x /\sqrt{2}} \zeta^{-1} +
                 e^{\lambda x /\sqrt{2}} \zeta^{-1} }
          { 2 \lambda } \\ & \\
  \ds\frac { e^{-\lambda^{\ss *} x /\sqrt{2}}\zeta^{*} +
                 e^{\lambda^{\ss *} x /\sqrt{2}}\zeta^{*} }
          { 2 \lambda^* } &
  \ds\frac { e^{-\lambda^{\ss *} x /\sqrt{2}} \zeta^{*} +
                 e^{\lambda x /\sqrt{2}} \zeta^{-1} }
          { \lambda +\lambda ^* }
 \end{array} \right),
    \end{equation}
whose value at the finite boundary, i.e., $M_0$, is
    \begin{equation} 
M_0(t) =
 \left( \begin {array}{cc}
  \ds \frac { \zeta^{-1} + \zeta^{*} }{ \lambda +\lambda ^* } &
  \ds \frac { \zeta^{-1} }{ \lambda } \\ & \\
  \ds \frac { \zeta^{*}}{ \lambda^* } &
  \ds \frac { \zeta^{*} + \zeta^{-1} }
          { \lambda +\lambda ^* }
 \end{array} \right) \equiv
\left( \begin {array}{cc} \mu & \nu \\ \rho & \mu \end{array} \right),
    \end{equation}
while its first-order spatial derivative at the same point is
    \begin{equation} 
\dot M_0(t) =
 \left( \begin {array}{cc}
  \ds \frac { - \lambda \zeta^{-1} + \lambda ^* \zeta^{*} }
            { \sqrt{2} ( \lambda +\lambda ^* ) }
  & 0 \\ 0 &
  \ds \frac { -\lambda ^*\zeta^{*} + \lambda \zeta^{-1} }
            { \sqrt{2} ( \lambda +\lambda ^* ) }
 \end{array} \right) \equiv
  \left( \begin {array}{cc} \epsilon & 0 \\ 0 & -\epsilon \end{array} \right),
    \end{equation}
where the dot denotes partial differentiation with respect to $x$.

In order to prove that the superposition of a soliton and its image has a
first-order  spatial derivative equal to zero at the \b, Eq.~\eqref{eq:Nss:syst}
may be written as
    \begin{equation}  \label{eq:stdr:uno}
M(x,t)u(x,t) = c \equiv \left( \begin {array}{c} 1 \\ 1 \end{array} \right),
    \end{equation}
and  differentiated with respect to $x$ to obtain
    \begin{equation}  \label{eq:stdr:dos}
\dot M(x,t)u(x,t) +  M(x,t) \dot u(x,t)=0.
    \end{equation}

Equation \eqref{eq:stdr:dos} yields
    \begin{equation} \label{eq:stdr:tres}
\dot u(x,t)= -M^{-1}(x,t) \dot M(x,t) M^{-1}(x,t) c,
    \end{equation}
while Eq.~\eqref{eq:stdr:uno} at the boundary yields
    \begin{equation} 
u_s = \frac {1}{|M_0| } (\mu-\nu), \qquad
\tilde u_s =  \frac {1}{|M_0| } (\mu-\rho),
    \end{equation}
which multiplied by $\dot M_0$ gives
    \begin{equation} 
\dot M_0 \left( \begin {array} {c} u_s \\ \tilde u_s \end{array} \right) =
  \frac{\epsilon}{|M_0|} \left( \begin {array} {c} \mu-\nu \\ -\mu+\rho
\end{array}
      \right).
    \end{equation}
Finally,  Eq.~\eqref{eq:stdr:tres} yields
    \begin{eqnarray} 
\dot u_s & = & - \frac {\epsilon}{|M_0|^2} ( \mu^2 - \nu \rho), \\
\dot {\tilde u_s} & = & - \frac {\epsilon}{|M_0|^2} ( -\mu^2 + \nu \rho),
    \end{eqnarray}
that clearly add to zero. Therefore, the superposition of a soliton and its
image with respect to a Neumann boundary satisfies homogeneous Neumann boundary
conditions at that boundary, and is a solution of the NLS in the quarterplane
subject to homogeneous Neumann boundary conditions at $x$=0.

Based on the results obtained in previous paragraphs, it may be thought that
an induction argument similar to the one employed in Section 4.1
could be used to demonstrate the validity of the method of images for the
$N$-soliton solution of the NLS equation in the quarterplane subject to
homogeneous Neumann boundary conditions at $x$=0. However, the tediousness of the
algebra involved has not allowed us as yet to present an elegant proof.
Following the same type of reasoning as in the proof for only  one soliton
and its image presented above and using  Matlab,
we have numerically  studied the method of images for a wide  range of parameters
with  up to ten  solitons. Based on these studies, we have numerical
evidence on the  validity of the method of images presented in this section for
quarterplane problems subject to homogeneous Neumann boundary conditions at
$x$=0.

In order to illustrate the method of images developed in this section, the
interaction of two  solitons of $A_1$=$V_1$=$A_2$=$-V_2$=$q$=1, $x_{01}$=0,
$x_{02}$=100, and $\phi_{01}$=$\phi_{02}$=0 is illustrated in
Figure~\ref{fg:2ss:neumann}. The results presented in
Figure~\ref{fg:2ss:neumann} simulate the interaction of a soliton with a
homogeneous Neumann boundary condition at $x$=50 in a quarterplane problem, and
indicate that the amplitude of both solitons increases as they approach $x$=50.
The results shown in Figure~\ref{fg:2ss:neumann} may be interpreted as either
the rebounding of a soliton from a Neumann boundary or the penetration of  two
colliding solitons. In either case, the solitons re-emerge from their interaction
with the  same amplitude and speed as those prior to their collision.

\section {The NLS Equation in the Finite Line}

\label{se:finiteline}

In this section, some numerical results of the NLS equation in finite lines
subject to homogeneous Dirichlet, Neumann and Robin boundary
conditions are presented and compared with those corresponding to periodic
boundary conditions and IVP problems. Special emphasis is placed on the
interaction of the solitons with the boundaries of the finite line. The results
presented in this section correspond, unless otherwise stated, to $q$=$A$=$V$=1,
$\phi_0$=$x_0$=0, $L$=50, $\Delta t$=0.01 and $\Delta x$=0.25.

\subsection {The Dirichlet Problem}

\label{se:dirichlet}

Some sample results corresponding to the solution in the NLS equation in the
finite line subject to homogeneous Dirichlet boundary conditions at both
boundaries, here referred to as the Dirichlet problem, are presented in
Figure~\ref{fg:dirichlet}. Note that, since the Dirichlet problem is invariant
under reflections in $x$, only the interaction of the soliton with the right
boundary is presented in this section.

Figure~\ref{fg:dirichlet} (top left)  indicates that, as the soliton approaches
the right boundary, its maximum amplitude increases due to the fact  that  the
soliton's amplitude is zero at the boundary, i.e., the soliton cannot penetrate
into the boundary. Furthermore, the soliton's velocity  decreases as the
soliton approaches the  right boundary. Note that, during the interaction of the
soliton with the boundary, a bump appears in the
soliton's tail away from the boundary. The amplitude  of the bump's minimum
decreases as the soliton approaches the right boundary.

When the soliton's velocity reaches a zero value, the bump's minimum also
reaches  a zero value, and the  soliton's maximum amplitude is largest.
Thereafter, the soliton rebounds from the right boundary and its largest
amplitude decreases  until both its speed and maximum amplitude recover the
values that they had  prior to the  collision of the soliton with the right
boundary.

The results presented in Figure~\ref{fg:dirichlet} (top left) are nearly
identical to those of Figure~\ref{fg:2ss:dirichlet} as it should be expected
since, for the finite problem considered here, the distance between the left and
right boundaries, i.e., $2L$=100, is very large compared with the soliton's
width and the interaction of the soliton with the right boundary in such a case
is expected to be nearly identical to that of the quarterplane problem presented
in Section~\ref{se:Dimages}.

The results presented in Figure~\ref{fg:dirichlet} (top right)  indicate that
the total mass is conserved during  the soliton propagation and interaction with
the right boundary as in the  IVP of the NLS equation. This result is consistent
with the integral of Eq.~\eqref{eq:evolmass} over the spatial domain.

The total momentum  illustrated in Figure~\ref{fg:dirichlet} (bottom left) is
constant when the soliton is far away from the boundaries. During the collision
between the \s\ and the \b, the total momentum experiences a change in sign
caused by the change in the direction of propagation after the soliton  rebounds
from the boundary. Figure~\ref{fg:dirichlet} (bottom left) also shows that the
total momentum changes  sign  smoothly during the  collision with the boundary.

The Hamiltonian or total energy is constant during the \s\ propagation, but it
suffers a great change as  the soliton collides  with the right boundary as
illustrated in Figure~\ref{fg:dirichlet} (bottom right) which indicates that the
total energy of the soliton first decreases as the soliton approaches the
boundary and then  increases as the soliton recedes from the boundary, until  it
recovers the constant value that it had prior to the interaction of the soliton
with the right boundary. The behaviour of the total momentum and total energy
described in previous  paragraphs is identical for all the collisions with
boundaries subject to homogeneous \D\ \bc s.

\subsection {The Neumann Problem}

\label{se:neumann}

The interaction of a soliton with a homogeneous Neumann boundary condition is
quite  different from that observed in the Dirichlet problem as illustrated in
Figure~\ref{fg:neumann} (top left). This figure shows that  the soliton
penetrates into the right boundary due to the zero  slope condition imposed
there. This penetration process is accompanied by an  increase in the soliton's
amplitude at the boundary and a decrease in the soliton's maximum  amplitude
away from the boundary; between these two relative maxima, the soliton's
amplitude exhibits a relative minimum.

Some time during the interaction of the soliton with the right boundary, the
maximum amplitude at the boundary becomes equal to that away from the boundary
which keeps on decreasing with a decreasing velocity. Later on, the largest
amplitude of the soliton occurs at  the boundary; therefore, the location of the
soliton's maximum amplitude undergoes a jump from the  interior of the domain to
the boundary, while the soliton's minimum amplitude keeps on decreasing as the
soliton approaches the right boundary.

When the soliton's minimum amplitude is
exactly zero, the soliton's maximum amplitude at the boundary is largest, and
the maximum amplitude away from the boundary is smallest and has a zero
velocity. At this moment, the soliton's maximum amplitudes at the boundary and
in the domain are twice and one half, respectively, the amplitude of
the soliton prior to the collision with the boundary.

As the soliton rebounds from the right boundary, the amplitude at the
boundary decreases while the maximum and minimum amplitudes away from the
boundary increase, the rebounding process is opposite to the collision one, and
the rebounding soliton recovers both the shape and the speed of the colliding
one.

It is important to note that, after the rebound from the boundary, the solitons
of both the \D\ and the \N\ problems move with  the same velocity and are
identical except for a phase difference of $\pi$ radians. Furthermore, since the
NLS equation subject to homogeneous  Neumann boundary conditions represents a
symmetric boundary value problem, i.e., the equation and the boundary conditions
are invariant under reflections in $x$, the interaction of the soliton with the
right boundary is identical to that with the left one. Therefore, a soliton
which undergoes two collisions one   with the right boundary followed by another
one with the left boundary will recover its original shape, velocity and phase
if both boundaries are subject to homogeneous Dirichlet or Neumann boundary
conditions. Moreover, after a  collision with the right boundary followed by
another one with the left boundary, there is no phase difference between  the
\D\ and  \N\ \s s.

The results presented in Figure~\ref{fg:neumann} (top left) are nearly identical
to those of Figure~\ref{fg:2ss:neumann} as it should be expected since, for the
finite line problem considered here, the distance between the left and right
boundaries, i.e., $2L$=100, is very large compared with the soliton's width and
the interaction of the soliton with the right boundary in such a case is
expected to be nearly identical to that of the quarterplane problem presented in
Section~\ref{se:Nimages}.

Figure~\ref{fg:neumann} (top right) indicates that the mass is nearly conserved
during  the soliton's propagation and interaction with the homogeneous Neumann
boundary condition  as in the  IVP of the NLS equation. According to
Eq.~\eqref{eq:evolmass}, the mass should be strictly conserved for the Neumann
problem; therefore, the non-constancy of the mass illustrated in
Figure~\ref{fg:neumann} (top right) is entirely due to small, numerical errors.

Figure~\ref{fg:neumann} (bottom) also shows that the total momentum is
constant when the soliton is far away from the boundaries and changes sign
during the soliton's interaction with the right boundary. This change is due to
the change in the soliton's velocity after the soliton  rebounds from the
boundary. Figure~\ref{fg:neumann} (bottom left) also shows that the total
momentum first   undergoes a small increase in  value at the beginning of the
collision process, and then rapidly changes its  sign reaching a  zero value at
the same time as that of the \D\ problem. The different slopes of
Figure~\ref{fg:dirichlet} (bottom left) and Figure~\ref{fg:neumann} (bottom
left) indicate that the sign change in the total momentum is  faster in the \N\
problem  than in the \D\ one.

Figure~\ref{fg:neumann} (bottom right) illustrates that the  total energy first
increases slightly  at the beginning of the collision process and then
decreases until it reaches a minimum value. Thereafter, it increases a
little bit until it  recovers the constant  value that it had prior to  the
collision with the boundary. Note that the slope and minimum value   of the
total energy are steeper and smaller, respectively, for the \N\ problem than
for the \D\ one.

The behaviour of the total momentum and the total energy described in previous
paragraphs is identical for all the collisions with boundaries subject to
homogeneous \N\ \bc s owing to the invariance of the Neumann problem under
mirror reflections in $x$.

\subsection {The Robin Problem}

\label{se:robin}

Since the NLS equation in the finite line subject to homogeneous Robin boundary
conditions at both boundaries is not a symmetric problem, the  interaction
between a \s\ with the right boundary is expected to be different from that with
the left one. For this reason, the interactions of the soliton with the right
and left boundaries are discussed separately in the next paragraphs.

Figure~\ref{fg:robin} (top left) shows the first collision of a soliton with the
right boundary for  $\gamma=1$ and indicates that, prior to the interaction with
the right  boundary, the  soliton behaves as in the IVP of the NLS equation,
i.e., it does not  notice the presence of the boundary.  At the beginning of the
collision with the  right boundary, the soliton's  amplitude at the boundary
has  a value approximately equal to half the sum of those of the \D\ and \N\
problems at the same boundary (cf. Sections 5.1 and 5.2). When  the \s\ is near
enough to the boundary,  its behaviour changes drastically.  First of all, the
soliton penetrates into the boundary, but it seems to be  retained in  a kind
of   boundary layer that  is formed at the right boundary and that seems to trap
and delay the soliton's   motion with respect to those observed in the \N\ and
\D\ problems.

Figure~\ref{fg:robin} (top left) also indicates that  the value of the slope
of the amplitude at the right boundary is  slightly smaller than that of the \D\
problem and that   the \s\ penetrates into the  boundary with a slightly larger
speed than that of the \N\ problem before  it
stops. This will illustrated in greater detail in
Section~\ref{se:solitonlocation}.

Figure~\ref{fg:robin} (top left) also shows that the largest amplitude of the
\s\  does not occur  at the boundary (compare with  the \N\ problem, cf.
Figure~\ref{fg:neumann} (top left)). The soliton does, however, show a minimum
and a maximum amplitude  near the right boundary. When the minimum amplitude
reaches a zero value, the maximum amplitude away from the boundary is largest and
the soliton velocity is nil. After the collision, the \s\ rebounds from the right
boundary undergoing a process similar, but opposite to that prior to the
collision, it appears to recover  its original shape and speed, and it exhibits
 a delay in its position with respect to the \P, \N\ and  \D\ problems caused by
the  boundary layer at the right boundary. This delay will illustrated in greater
detail in  Section~\ref{se:solitonlocation}.

Figure~\ref{fg:robin} (top right)  illustrates the collision that the soliton
undergoes with the left boundary after having interacted previouly with the
right one, and indicates that, at the beginning of the collision, the \s\ seems
to behave as in the collision with the right  boundary. However, the soliton does
not seem to penetrate into the left  boundary; rather, a hump is formed ahead of
the soliton, and this hump  penetrates into the boundary.

The hump has a similar but opposite slope to that of the  \D\  problem, and
joins the soliton's maximum amplitude with  a minimum of very large curvature.
During the  collision, the hump's amplitude increases, the curvature of the
minimum  amplitude also increases, the soliton's maximum amplitude does not
reach as high values as in the collision with the right boundary but is larger
than the hump's amplitude, and another minimum of decreasing amplitude appears
in the soliton's tail. The appearance of both the hump and the minimum near to
the left boundary may be caused by the boundary layer formed there which does
not allow the soliton to penetrate into the boundary.

When the minimum amplitude in the soliton's tail reaches a zero value, the
maximum amplitude of the soliton, the hump's amplitude and the minimum
amplitude near the boundary are largest, whereas the soliton's velocity is nil.

In the rebounding process, the \s\ appears to show an opposite behaviour to that
prior to the collision; however, the boundary layer at the left boundary traps
the soliton's tail in such a manner that,  after the collision, the \s\
re-emerges with a little hump localized at the left  boundary whose amplitude
remains nearly constant as the soliton keeps on propagating  towards the right
boundary.

As indicated in Figures~\ref{fg:dirichlet}--\ref{fg:robin}, the location of
the soliton's maximum amplitude in the Robin problem is delayed when it collides
with the right  boundary with respect to those  corresponding to the \D\ and \N\
problems. However, after the collision  with the left boundary, no delay with
respect to the solitons observed in the \N\ and \D\ problems is   observed.
This will be illustrated in greater detail in
Section~\ref{se:solitonlocation}.

In order to examine further interactions of the soliton with the boundaries,
the amplitudes at the right and left boundaries are presented in
Figure~\ref{fg:robin} (bottom left) and Figure~\ref{fg:robin}
(bottom right), respectively, as functions of time for $\gamma=1$ and
$\gamma=-1$. For $\gamma=1$, the amplitude at the right boundary is zero except
when the soliton collides with that boundary  (cf. Figure~\ref{fg:robin}
(bottom left)). The amplitude at the left boundary (cf. Figure~\ref{fg:robin}
(bottom right)) clearly indicates the formation of a hump in the first
collision with that boundary. The hump's amplitude increases as the soliton
collides with the left boundary until about $t=900$, it then decreases
until about $t=1800$. Further collisions with the left boundary indicate
that the hump's dynamics is periodic with a period of about nine
collisions with that boundary. This periodic behaviour is studied in
greater detail in Section~\ref{se:phasediagrams}.

Figure~\ref{fg:robin}
(bottom right) also indicates that the hump's amplitude does not change as the
soliton propagates from one boundary to the other one and when the soliton
collides with the right boundary. Furthermore, the results presented in  Figure~\ref{fg:robin}
(bottom) are qualitatively and quantitatively
different from  those observed in the Dirichlet and Neumann problems
(cf. Figures~\ref{fg:dirichlet} and~\ref{fg:neumann}) where
subsequent collisions with the boundaries preserve the soliton's original shape
and velocity, and are due to the lack of invariance of the Robin
problem under mirror reflections in $x$.

When  $\gamma=-1$, no hump appears on the left boundary; however, a hump is
formed at the right boundary due to the asymmetry of the homogeneous Robin
boundary conditions. This is  illustrated in Figure~\ref{fg:robin} (bottom) which
shows  that the dynamics and period of the hump at the right boundary for
negative values of $\gamma$ are exactly the same as those observed in the left
hump for positive values of $\gamma$,  except for the numerical errors
introduced by the discretization  which cause  slight variations in the   hump's
amplitude.

For values of $\gamma$ different from $1$ and $-1$, the collision of the soliton
with the boundaries is similar to that of the \N\ problem for $|\gamma|\gg 1$ or
the  \D\ one for $|\gamma| \ll 1$, as it should be expected (cf.
Eq.~\eqref{eq:nse:r}). For values of $|\gamma|= \OL(1)$, the interaction of the
soliton with the Robin boundaries is similar to that for $|\gamma|= 1$.

Numerical experiments not included here indicate that the  amplitude and
frequency of the hump's recurrent phenomena, the penetration of the
 soliton into the boundaries, the phase shift introduced in the collisions with
the boundaries and  the delay in the location of the maximum amplitude that  the
soliton experiences upon colliding with a Robin boundary depend strongly  on
$\gamma$ and are largest for $|\gamma|=1$. These numerical experiments also
indicate that the dynamics of  the left and right humps are nearly periodic with
periods of about $9$ and $2$ collisions for  $\Delta x=0.25$ and $0.0625$,
respectively, and that the humps exhibit some kind of noise or oscillations, the
magnitude of which decreases as the time step is decreased. However, the trends
shown in Figure~\ref{fg:robin} are qualitatively independent of the spatial grid size
employed in the calculations.

Figures~\ref{fg:robin:mass}--\ref{fg:robin:energy} show a detailed view of the
mass, total momentum and total energy for several values of  $\gamma$ as the
soliton collides with the right and left boundaries. Figure~\ref{fg:robin:mass}
indicates that the mass  suffers some  changes, but it recovers the value that
it had prior to the collision, after the soliton rebounds from the right and
left  boundaries. For positive values of $\gamma$, the collisions with the right
and left boundaries produce a peak and a valley  in the mass, respectively,
whereas,  for negative values of $\gamma$,  the opposite behaviour is observed.
The magnitudes of the peak and valley are  largest for $|\gamma| \approx 2$  and
decrease monotonically  as $|\gamma|$ tends to either infinity or zero.

Figure~\ref{fg:robin:momentum} shows that the total momentum maintains a
constant value during the \s\ propagation and  suffers a change in sign when the
soliton collides with the right (Figure~\ref{fg:robin:momentum} (left)) and left
(Figure~\ref{fg:robin:momentum} (right)) boundaries due to the change in the
direction  of the \s\ velocity  after  rebound. For  $|\gamma| > 2$, the total
momentum has a similar behaviour to that observed in the \N\ problem
(cf. Figure~\ref{fg:neumann}), whereas, for other values of $\gamma$, its
behaviour is similar to that of the Dirichlet problem
(cf. Figure~\ref{fg:dirichlet}). Figure~\ref{fg:robin:momentum} also shows that
the largest value of the time derivative of the total momentum increases as
$|\gamma|$ is increased, and that topological changes occur near the value
$|\gamma| \approx 2$.

Figure~\ref{fg:robin:energy} indicates that the total energy is constant during
the \s\ propagation, and that the changes that it undergoes when the soliton
collides with the boundaries depend on the value of $\gamma$.  The total energy
recovers the value that it had prior to the collision, after the soliton
rebounds from the boundary. For positive values of $\gamma$, the total energy
exhibits a  valley at the right boundary which is deeper than those observed in
the \D\ and \N\  problems. The smallest magnitude of this valley occurs for a
value of $\gamma$ approximately  equal to 2. For   more \N\ problems, i.e., for
$\gamma \geq 10$, the total energy  exhibits a valley with two small peaks
surrounding it when the soliton collides  with the right boundary as indicated in
Figure~\ref{fg:neumann}. These peaks cannot be observed in Figure~\ref{fg:robin:energy}
(left) because of the large scale employed to capture the smallest value of the
energy.

In the next collision which occurs at the left  boundary, the total
energy  exhibits a peak, rather than a valley as indicated in Figure~\ref{fg:robin:energy}
(right),  except for the more \N\ problems, i.e., $\gamma \geq 10$, not
illustrated here for which it shows  a valley surrounded by two  small peaks,
i.e., it exhibits a  similar shape to the one observed in the first collision
with the rigth boundary.

The results presented in Sections~\ref{se:dirichlet} and~\ref{se:neumann}
indicate that the Dirichlet and Neumann problems which are  limiting cases of
the Robin one, have  identical dynamics at the left and right boundaries.
However, the  \R\ problem with $|\gamma| \leq 10$ yields  some asymmetry caused
by the lack of invariance  of the Robin boundary conditions under reflections in
$x$.

\subsection{The Soliton's Location and Maximum Amplitude}

\label{se:solitonlocation}

Figure~\ref{fg:location:PND} (left) shows the maximum amplitude of the soliton as
a function of time for the periodic, Dirichlet and Neumann problems and indicates
that, for the \P\ problem, the soliton's maximum amplitude is nearly one except
for the small  oscillations caused by the finiteness of the spatial grid
employed in the calculations. For the  \D\ problem, the soliton's maximum
amplitude grows smoothly during the  collision of the soliton with the right
boundary. Figure~\ref{fg:location:PND} (left) also shows that the dynamics of the
soliton's maximum amplitude in the Neumann problem is  more complex than those
of the periodic and Dirichlet ones. In particular, the soliton's maximum
amplitude exhibits  an initial  decrease followed by a rapid and  large
increase. However, as indicated in Section~\ref{se:neumann}, this  complex
dynamics is associated with the jump that the location of the maximum  amplitude
undergoes as the soliton collides with the boundary.

Figure~\ref{fg:location:PND} (right) illustrates the location of the soliton's
maximum amplitude and the solitons's crossing of the right boundary in the
periodic problem although the left boundary is not shown in the figure. Figure~\ref{fg:location:PND} (right)
also illustrates that the soliton does not penetrate into the boundary in the
\D\ problem, and that the location of the soliton's maximum amplitude jumps
from the interior of the domain to the boundary and vice versa in the \N\
problem.  Figure~\ref{fg:location:PND} (right) also indicates that the velocity
of the soliton is constant when the soliton is sufficiently far away from the
boundaries and that no delay between the \D\ and \N\ solitons is observed after
their collisions with the right boundary.

The soliton's maximum amplitude and its
location for the Robin problem are illustrated in Figure~\ref{fg:location:robin}
as a function of time for several values of $\gamma$.  Except for  the
discontinuous change in the location of the maximum  amplitude for the more \N\
problems, the results presented in Figure~\ref{fg:location:robin} (top) are
similar to those presented  previously. Figure~\ref{fg:location:robin} (bottom)
indicates  clearly that the delay in the location  of the soliton's maximum
amplitude   after  it collides with the boundaries depends on the value of
$\gamma$,  and that  the delay produced in a boundary is compensated for by the
collision of  the soliton with the other  boundary.

It must be pointed out that, due to numerical errors, the numerically
determined speed of the soliton prior to its collision with the
boundaries is slightly smaller than that of the IVP of the NLS equation, i.e.,
$V$=1. Moreover, these numerical errors and the finite-order approximation used
to discretize both the boundary  conditions and the NLS equation in the finite
line cause a slight difference in the location of the soliton's maximum
amplitude in long time  integration, e.g., more than ten collisions with each
boundary, for the Dirichlet and Neumann problems. Furthermore, the results shown
in Figures~\ref{fg:dirichlet}--\ref{fg:robin} indicate the presence of
small-amplitude, background noise or radiation in the tails of the solitons,
which is larger for the Neumann problem than for the Dirichlet one. The
magnitude of this noise or radiation decreases as the grid is refined, and
cannot be clearly appreciated in Figures~\ref{fg:dirichlet}--\ref{fg:robin} due
to the scale of these figures.

\subsection{Phase Diagrams, Recurrence and Nonlinear Dynamical Effects}

\label{se:phasediagrams}

In the results presented in previous sections, a recurrent behaviour has been
observed due to  the collision of the soliton with the boundaries. In order to
analyze this recurrent phenomenon in more detail, the phase  diagrams
corresponding to the maximum amplitude of the soliton and to the amplitudes at
the left and right boundaries are studied in this section. A  second-order
accurate, finite difference method has been employed to determine  the phase
diagram velocity.

For  the \P\ problem, the soliton's propagation is not affected by the \bs;
therefore,  its maximum amplitude is constant, and the corresponding phase
diagram  is a  fixed point. Furthermore, the dynamics of the amplitude at the
left and right   boundaries is the same as that at  any point  in the finite
interval. For this  reason, the phase diagram will be only presented at the
point $x$=0.

The amplitude and the time derivative of the amplitude for the exact 1-soliton
solution of the IVP of the NLS equation (cf. Eq.~\eqref{eq:1ss}) are
    \begin{equation} \label{eq:phidef}
\psi(t) \equiv |u(0,t)| = A \sech (\mu),
    \qquad \mu = -\frac{A}{\sqrt{2}} (x_0+Vt),
    \end{equation}
    \begin{equation} \label{eq:toderive}
\dpar{}{t} \psi(t) = - \mu_t \psi(t) \tanh (\mu),
    \end{equation}
which yield the following  phase diagram
 \[
\left\{
  \begin{array}{l}
    \dot \nu  =  \frac{A^2 V^2}{2} \left( \psi - \frac{2}{A^2} \psi^3 \right)
\\  \\
    \dot \psi =  \nu,
     \end{array}
\right.
 \]
whose fixed points are $(\psi,\nu)$ = (0,0) =
$(A/\sqrt {2}$,$0)$ = $(-A/\sqrt {2}$,$0)$. A linear stability analysis
indicates that the point (0,0) is a saddle point, while the other two are focal
points. The third fixed point, however, lacks physical meaning.

Figure~\ref{fg:phase:PND} (top) shows that the phase diagram corresponding to the
exact 1-soliton solution of the IVP or to the periodic problem of the NLS
equation resembles that of the unforced Duffing
equation~\cite{Thom86}.  This is not surprising since  Ablowitz et
al.~\cite{Ablo91b}  have  indicated that the Duffing equation can be derived
from the NLS equation using a method  of separation of variables. Therefore, a
mass-string system with a cubic nonlinear stiffness may be a good model of the
dynamics of the \s's amplitude for the IVP and periodic problem of the NLS
equation.

Figure~\ref{fg:phase:PND} (top) also shows that the phase diagram of the
soliton's amplitude at the  right boundary is a fixed point located at the
origin of the phase plane for the \D\ problem, since the \s's amplitude is
always equal to zero at the boundaries.

The dynamics of the amplitude at the left and right boundaries is identical for
the \N\ problem, has been evaluated numerically, and is also illustrated in
Figure~\ref{fg:phase:PND} (top). This figure indicates that
the phase diagram for the Neumann problem is similar to that of the \P\
boundary conditions; however, its largest amplitude and velocity are about two
and twenty, respectively, times larger than those of the \P\  problem for
$\Delta x=0.25$, and about twice and the same,  respectively, as those of the
\P\  problem for $\Delta x=0.0625$ (not shown here). The large  velocity
differences in the phase diagrams  as the
grid size is decreased, clearly indicate  the coupling between the temporal and
spatial phenomena associated with the  soliton propagation, while the magnitude
of these velocities is mainly due to  the  steepening of the soliton as it
collides with the  \bs.

The phase diagram of the soliton's maximum amplitude  for the \P, \D\ and \N\
problems are shown in Figure~\ref{fg:phase:PND} (bottom). The phase diagram for
the \P\ problem is the fixed point $(1,0)$, while that for the \D\ one is nearly
elliptical except for the oscillations caused by the spatial grid size used in
the calculations. Figure~\ref{fg:phase:PND} (bottom) also indicates that the
phase diagram of the soliton's maximum amplitude for the \N\ problem is
topologically very  different from those of the \P\ and \D\ ones. In particular,
the phase diagram for the \N\ problem  clearly indicates the large  velocities
associated with the jump that the location of the soliton's maximum amplitude
undergoes from the interior of the domain to the boundary as the soliton
approaches the boundary and from the boundary to the interior of the domain as
the soliton rebounds from the boundary (cf. Section~\ref{se:neumann}).
Figure~\ref{fg:phase:PND} (bottom) also illustrates the smooth behaviour of the
phase diagram as the soliton propagates from one boundary to the other one, and
resembles those observed in relaxation oscillations (cf., e.g.,~\cite{Thom86}).
Figure~\ref{fg:phase:PND} (bottom) also shows the  background noise or radiation
introduced by the discretization, which is characterized by a fat point in the
phase diagram for the periodic problem.

The closed phase diagrams for  the \D\ and \N\ problems confirm the   recurrent
behaviour of  solitons in finite lines subject to homogeneous \D\ and \N\
boundary conditions at both boundaries.

Figure~\ref{fg:phase:robin1} shows the phase diagrams corresponding to  the
amplitude at the left and right boundaries for  several values of $\gamma$.
These diagrams are similar to  that corresponding  to \P\ boundary
conditions (cf. Figure~\ref{fg:phase:PND}), and indicate that  the maximum
amplitude at the boundaries increases  monotonically from zero to 2 as
$|\gamma|$ is varied from zero to infinity and, is almost linear  for $\gamma
\ll 1$. The maximum velocity of  the amplitude at the boundary points is about
one half  the maximum amplitude there for all  values of $\gamma$.

As shown in Section~\ref{se:robin}, the lack of invariance of the \R\ problem
under mirror reflections in $x$ results in the formation of a boundary layer
near to the left boundary for $\gamma = \OL (1)$. This boundary layer traps the
soliton and creates a hump at the left boundary. For negative values of
$|\gamma| = \OL (1)$, a hump is formed at the right boundary, and this hump
exhibits exactly the same dynamics as that of the left one as indicated in
Figure~\ref{fg:robin}. For this reason, only the dynamics of the  hump formed
at the left boundary will be discussed in the following paragraphs.

The phase diagrams corresponding to the amplitudes at the left and right
boundaries are presented in Figure~\ref{fg:phase:robin2} (left) and
Figure~\ref{fg:phase:robin2} (right), respectively. Figure~\ref{fg:phase:robin2}
(right) indicates that the phase diagram of  the amplitude at the right boundary
coincides with that of the \P\ problem, whereas that of the amplitude at the
left boundary closes upon itself after nine collision of the soliton with that
boundary, for $\gamma=1$. For $\gamma=-1$, no hump is formed at the left boundary
and the phase diagram of the amplitude at that boundary coincides with that
of the \P\ problem, whereas the phase diagram of the amplitude at the right
boundary closes upon itself after nine collisions of the soliton with that
boundary. As shown in Section~\ref{se:robin}, the
nonlinear dynamics of the hump is very sensitive to the spatial grid size and to
the value of $\gamma$;  small changes  in $\gamma$ or in the spatial grid size
result in large changes in  both the  period and the maximum amplitude of the
hump's oscillations.

The phase diagrams of the boundary amplitude shown above indicate that the
Duffing equation may be used to study the dynamics of solitons in finite lines
with homogeneous Robin boundary conditions if some forcing is introduced in that
equation to represent, in some manner, the boundary layers that appear at the
boundaries of finite lines.

The  phase diagram corresponding to the  soliton's maximum amplitude in a finite
line with homogeneous \R\ \bc s  is presented in  Figures~\ref{fg:lobes:robin1}
and~\ref{fg:lobes:robin2} for several values of $\gamma$.  For
$\gamma=1.0$, the phase diagram presents two lobes (cf.
Figure~\ref{fg:lobes:robin1}). The lobe of  larger amplitude, here referred to
as the first lobe,  is caused by the collision of the soliton with the  right
boundary where no hump is formed, and the amplitude of this lobe is larger than
the maximum amplitude  of the  Dirichlet problem. The other, smaller lobe, here
referred to as the second lobe, is caused by the collision of the soliton with
the left boundary  where a hump which traps and delays the soliton is formed.
The maximum amplitude of the second lobe is smaller than the maximum amplitude
of the  Dirichlet problem. The amplitude of the second lobe decreases whereas
that of the first one increases as $\gamma$ is increased from one.

For a value of  $\gamma \approx 1.1$, the results presented in
Figure~\ref{fg:lobes:robin2} indicate that the second lobe  has a different
shape than the elliptical one described in Figure~\ref{fg:lobes:robin1} and
exhibits an additional lobe, here referred to as third lobe, which is
characterized by very high velocities due to the fact that the hump's amplitude
becomes equal to or exceeds the largest amplitude of the soliton away from the
boundary (cf. Section~\ref{se:neumann}). When this occurs the velocity shown in
the phase diagram increases rapidly as indicated in the Neumann problem. The
amplitudes of the first and  second lobes increase and decrease, respectively,
while that of the third lobe increases quite rapidly, as $\gamma$ is increased.
For $\gamma \approx 2$, the second lobe has nearly disappeared, while the third
lobe  is as large as the  first one.

For $\gamma > 2$, the third lobe has approximately the shape of
half an ellipse  connected with a very steep, nearly straight line to a nearly
triangular shape (cf. Figure~\ref{fg:lobes:robin1}). For still larger values of
$\gamma$, the amplitude of the third lobe is larger than that of the first one,
which, in turn, acquires a similar shape to that of the third lobe
(cf. Figure~\ref{fg:lobes:robin1}). For $\gamma
= \infty$, i.e., for \N\ boundary conditions, the first and third lobes are
identical, indicating that these lobes collapse into only one as  $\gamma$ tends
to infinity. Since  the NLS equation subject to \N\ \bc s is a symmetric problem
in $x$, the qualitative changes observed in the phase diagrams presented in
Figures~\ref{fg:lobes:robin1} and~\ref{fg:lobes:robin2}   are related to  the
loss of symmetry of that equation in the finite line as $\gamma$ is varied.

\subsection{Area of the Phase Diagrams}

\label{se:areas}

An indication of the topological changes that occur in the phase diagrams as
$\gamma$ is varied can be obtained by evaluating the areas enclosed by them.
This evaluation also gives information on the conservation properties of the
numerical scheme employed in the calculations. In this section, the areas of
the phase diagrams presented in the previous section are calculated.

In order to evaluate the areas of the phase diagrams, it is necessary to account
for the numerically-induced  noise or background radiation which, in spite of
its small amplitude, is responsible  for the fat points observed in the phase
diagrams, e.g., Figure~\ref{fg:phase:PND} (bottom),~\ref{fg:lobes:robin1}
and~\ref{fg:lobes:robin2}. If the background radiation were used in the
evaluation of these areas,  noisy results would be obtained. It is, however,
possible to  reduce this noise by filtering the soliton's maximum amplitude and
the amplitude at the boundary points in such a manner that the filtered  phase
diagram preserves the most important features, i.e., those associated with the
collisions of the soliton with the boundaries, of the unfiltered diagrams.  The
filter used in this paper employs an amplitude threshold to locate the relative
minima closest to both the threshold amplitude and the amplitude at the boundary
or the soliton's maximum amplitude. Once these  minima have been found, only the
part of the maximum amplitude and those at the boundaries between these minima
are used to draw the filtered phase diagrams whose areas are evaluated
numerically by means of Simpson's quadrature rule.

The areas of the phase diagrams corresponding to the amplitude at the left and
right boundaries are illustrated in Figure~\ref{fg:areas} (top) as functions
of the number of collisions of the soliton with the boundaries for several values
of $\gamma$. This figure indicates that, for $\gamma \gg 0$, the areas of the
phase diagrams decrease less than the $0.5\%$ after ten collisions with each
boundary, indicating that dissipative effects are small. Figure~\ref{fg:areas}
(top) also indicates that the areas of the phase diagrams decrease in an
oscillatory manner and that the largest amplitude of these  oscillations is
observed for $\gamma= \OL (1)$. These oscillations seem to be introduced by both
the background radiation and the filter used to evaluate the areas, and are
substantially reduced as the spatial grid size is reduced. Note that the
background radiation decreases as $\Delta x$ is decreased.

For $\gamma \ll 1$, the areas of the phase diagrams increase slightly with the
number of collisions of the soliton with the boundaries because the magnitude of
the  background radiation is comparable to or larger than the amplitude at the
boundaries. Note that $\gamma=0$ corresponds to the \D\ problem for which the
boundary amplitude is exactly zero.

Figure~\ref{fg:areas} (top) also shows that the areas of the phase diagrams for
the boundary points increase monotonically from zero for the \D\ problem  to
about  2.5 for the  \N\ one.

The area of the  phase diagram corresponding to the soliton's maximum amplitude
is  also presented in Figure~\ref{fg:areas} (bottom), which indicates the
effects of the background radiation for small values of $\gamma$. For very
large values of $\gamma$,  the area of the phase diagrams varies in an
irregular manner because of the errors introduced by the filtering technique
and, most importantly, by the fact that  the location of the soliton's maximum
amplitude  jumps discontinously from the interior of the domain to the boundary,
and vice versa. For $\gamma= \OL (1)$, the area of the phase diagrams  clearly
shows the  lack of invariance of the  \R\ problem under mirror reflections in
$x$. For $\gamma \approx 1$,  the area of the phase diagram
is larger when the soliton collides with the right boundary than when it
collides with the left one in agreement with the lobes presented in
Figures~\ref{fg:lobes:robin1} and~\ref{fg:lobes:robin2}, and it exhibits some
oscillations,  the amplitude of which is largest for $\gamma \approx 1.7$ but
does not exceed $0.4\%$.

Figure~\ref{fg:pitch} shows the area of the phase diagrams corresponding to the
amplitude at the left and right boundaries (top) and to the soliton's maximum
amplitude (bottom) as functions of
$\gamma$. Note that  Figure~\ref{fg:pitch} (top left) and Figure~\ref{fg:pitch} (top
right) correspond to the sixth collision of the soliton with the left and
right, respectively, boundaries.

The results presented in Figure~\ref{fg:pitch} (top) indicate that the area of
the phase diagram corresponding to the amplitude at the left boundary is
nearly identical to that coresponding to the amplitude at the right one
despite the lack of invariance of the \R\ problem under mirror reflections in
$x$. The results presented in Figure~\ref{fg:pitch} (top) must be interchanged
for negative values of $\gamma$.

Figure~\ref{fg:pitch} (bottom) shows the areas of the phase diagram
corresponding to the maximum amplitude for collisions of the soliton with the
left and right boundaries as a function of $\gamma$. Figure~\ref{fg:pitch} (bottom
right) indicates that the area of the phase diagram for collisions of the
soliton with the right boundary is larger than that for collisions with the
left one, in agreement with the results presented in Figure~\ref{fg:areas}.
The largest differences  between the areas of the phase diagrams corresponding
to collisions of the soliton with the left and right boundaries (cf. Figure~\ref{fg:pitch}
(bottom)) are observed for values of $\gamma$ between 0.6 and around
1.7, and are due to the formation of a boundary layer near the left
boundary which does not allow the soliton to penetrate into that
boundary (cf. Section~\ref{se:robin}) and the consequent formation of
the first and second lobes shown in Figures~\ref{fg:lobes:robin1}
and~~\ref{fg:lobes:robin2}.

\section{Conclusions}

The NLS equation subject to  homogeneous \D, \N\ and \R\ \bc s in the finite
line has been studied numerically by means of a second-order accurate
Crank-Nicolson method, employing as initial condition the \eNss\ truncated to
the finite line and translated in such a manner so as to avoid mathematical
incompatibilities with the boundary conditions.

A method of images similar to the one used in potential theory has also been
developed to study the NLS equation in the quarterplane subject to homogeneous
\D\ or \N\ boundary conditions at the finite boundary. It has been shown that
the interaction of a soliton in quarterplane problems with homogeneous \D\
boundary conditions is equivalent to the interaction of two solitons of equal
and opposite amplitude and velocity and identical phase placed symetrically with
respect to the finite boundary, whereas the interaction of a soliton in
quarterplane problems with homogeneous \N\ boundary conditions is equivalent to
the interaction of two solitons of equal and opposite velocity and identical
phase and amplitude placed symetrically with respect to the finite  boundary.

Numerical simulations of the \s's interaction with the \D\ and \N\ \bc s in
finite line problems indicate that the \s\ rebounds from the \bs\ recovering its
shape except for a phase shift and a numerically induced  background radiation
whose magnitude decreases as the mesh size is decreased.  The \s's amplitude
increases and a bump is formed in the  \s's tail during the collision with the
\D\ boundary. The rebounding process starts when the bump's minimum reaches a
zero amplitude, the \s's amplitude reaches its largest value and the \s's
velocity is exactly zero.

The collision of a soliton with a \N\ boundary in finite lines is qualitatively
quite different from that with a \D\ one since the amplitude at the boundary
increases from zero until a maximum, larger than the soliton's amplitude in the
interior of the domain, is reached. Later on, the amplitude at the boundary
decreases while the maximum amplitude in the interior increases. As a
consequence, the location of the soliton's maximum amplitude behaves
discontinuously since it jumps from the interior to the boundary as the soliton
approaches it and from the boundary to the interior as the soliton rebounds from
the boundary. It has also been shown that the collision process with the left
and right boundaries   are identical owing to the symmetry of \ibvp s of
the NLS equation subject to homogeneous Dirichlet or Neumann boundary conditions
in finite lines.

The homogeneous \R\ \ibvp\ of the NLS equation in the finite line is not
invariant under mirror reflections in the spatial coordinate. As a consequence,
the collisions between the \s\ and the left and right \bs\ have different
qualitative behaviour.  For positive values of the  coefficient that defines the
\R\ \bc s, the \s\ penetrates into the right \b\ reaching a  large amplitude
there, and a bump is formed in its tail. This bump behaves  as those of the \D\
and \N\ problems, but its motion is delayed by some kind of boundary layer that
traps the \s. After the collision with the right boundary, the rebounding \s\ is
further delayed with  respect to those of the \D\ and \N\ problems.  In the
collision with the left \b, a boundary layer which does not allow for the
penetration of the \s\ into the \b, and a hump are formed. After the soliton
rebounds from the left boundary, the hump's amplitude is almost constant and
subsequent collisions with the right \b\ are not affected by this hump. However,
subsequent collisions with the left \b\  increase  the hump's amplitude until it
reaches a maximum, beyond which, further collisions with this boundary cause a
decrease in the hump's amplitude. The magnitude of the period and maximum
amplitude of the periodic hump's behaviour  depend on the mesh size and are
almost independent of the time step used in the numerical calculations.  For
negative values of $\gamma$, a hump is formed near to the right boundary.

For the homogeneous \N\ and \D\ \bc s, the mass of the soliton is exactly
conserved  in the finite line, while the total momentum changes sign during the
collision process but it  recovers the value that it had prior to the collision
with the boundary. The total energy  decreases and increases as the soliton
approaches and recedes from, respectively, the boundary, indicating a transfer of
energy between the \s\ and the \b.

For the  homogeneous \R\ \bc s, the mass of the soliton is conserved between
collisions, but it increases and decreases as the soliton collides with the
right and  left boundaries, respectively. The total momentum changes sign
during  the collision of the soliton with the boundary, and its largest value
takes place  for $|\gamma| \approx 2$. The total energy exhibits a valley and a
peak for collisions with the right and  the left boundaries, respectively,
except for $|\gamma| \gg 10$ for which it exhibits only a valley  in both \bs.

The phase diagrams of the \s\ amplitude at the boundary points and for the
soliton's maximum amplitude indicate the recurrent behaviour of the collisions of
the solitons with the boundaries. The phase diagram  has been calculated
analytically for periodic boundary conditions and exhibits similar trends to
those of the cubic Duffing equation. The phase diagram of the maximum amplitude
for the \D\ \bc s has an almost elliptical shape and exhibits some oscillations.
For \N\ \bc s, the phase diagrams of the  amplitude at the \b\ points have the
same shape as in the \P\ case, while that of the maximum amplitude is
topologically very different from those of the \D\ case and shows large
velocities associated with the jump that the location of the \s's maximum
amplitude  experiences  upon the collision of the soliton with the boundaries.

For \R\ \bc s, the phase diagrams for the amplitude at the \bs\ are similar to
those of the \P\ case, but the maximum amplitude reached at the \b\  depends
monotonically on $|\gamma|$ and it is 0 for the \D\ case and 2 for the \N\  one.
For $|\gamma|$=$\OL(1)$, the hump that appears at one of the \bs\ causes  that
the phase diagram  closes upon itself after various collisions with this \b. The
phase diagram corresponding to the maximum amplitude depends strongly on the
value of $\gamma$.

The areas of the phase diagrams for both the soliton's maximum amplitude and the
amplitudes at the boundaries have been evaluated numerically by means of a
filter in order to  eliminate some of the background radiation.  The area of the
phase diagram of the amplitude at the \bs\ decreases slightly in an oscillatory
manner due to the dissipative effects introduced by the numerical  method,
whereas the area of the  phase diagram for the soliton's maximum amplitude
clearly shows the non-invariance of the homogenous Robin \bc s under mirror
reflections in the spatial coordinate which results in larger areas for the
collisions  of the soliton with the right \b\ than for those with the left one.

\section*{Acknowledgments}

This research was supported by the Spanish D.G.I.C.Y.T. under Project  no.
PB91-0767. The second author (F.R.V.) has a fellowship from the
Programa Sectorial de Formaci\'on de Profesorado Universitario y Personal
Investigador, Subprograma de Formaci\'on de Investigadores "Promoci\'on General
del Conocimiento", from the Ministerio de Educaci\'on y Ciencia of Spain.



\newcommand{\bookref}[6]{#1, {\em{#2}}, #3 (#6).}

\newcommand{\paperref}[6]{#1, #2, {\em {#3}} {\bf{#4}}, #5 (#6).}

\newcommand{\paperrefiss}[7]{#1, #2, {\em {#3}} {\bf{#4}}(#7) #5 (#6).}

\newcommand{\procref}[7]{#1, #2, In {\em{#3}}, (Edited by #4), pp. #6, #5,
(#7).}

\newcommand{\procrefvol}[9]{#1, #2, In {\em {#3}}, #4, (Edited by #5), #6,
vol. #7, pp. #8 (#9).}

\newcommand{\procrefv}[8]{#1, #2, in {\em #3}, (Edited by #4), {\em #5} vol. #6,
pp. #7 (#8).}

\clearpage

\section*{Figure Captions}

\begin{enumerate}

\item {Amplitude, $|u|$, of two interacting solitons of
$A_1$ = $V_1$ = $-A_2$ = $-V_2$ = $q$ = 1, $x_{01}$=0, $x_{02}$=100, and
$\phi_{01}$=$\phi_{02}$=0 as a function of space and time calculated using the
\eNss\ of the NLS equation. Note that $u(50,t)=0$. }
 \label{fg:2ss:dirichlet}

\item {Amplitude, $|u|$, of two interacting solitons of
$A_1$ = $V_1$ = $A_2$ = $-V_2$ = $q$ = 1, $x_{01}$=0, $x_{02}$=100, and
$\phi_{01}$=$\phi_{02}$=0 calculated using the \eNss\ of the NLS equation. Note
that $u_{x}(50,t)=0$.}
 \label{fg:2ss:neumann}

\item {Amplitude, $|u|$, (top left), mass (top right), momentum (bottom left)
and  energy (bottom right)   of a
soliton colliding with   a homogeneous \D\ \bc\ located at
$x$=50.}
 \label{fg:dirichlet}

\item {Amplitude, $|u|$, (top left), mass (top right), momentum (bottom left)
and  energy (bottom right)   of a
soliton colliding with   a homogeneous \N\ \bc\ located at
$x$=50.}
 \label{fg:neumann}

\item {Amplitude, $|u|$, of a soliton colliding with  the right (top left) and
the left (top right) boundaries of a finite line  subject to homogeneous Robin
boundary conditions for $\gamma$=1, and amplitude at the right (bottom left)
and left (bottom right) boundaries  for $\gamma$=1 (continuous line) and
$-1$ (dashed line).}
 \label{fg:robin}

\item {Mass of a soliton colliding with the right (left) and left (right)
boundaries of a finite line subject to homogeneous Robin boundary conditions
for $\gamma$=10~(continuous line), 2~(dashed line), 1~(dotted-dashed line) and
0.1~(dotted line).}
 \label{fg:robin:mass}

\item {Momentum of a soliton colliding with the right (left) and left (right)
boundaries of a finite line subject to homogeneous Robin boundary conditions
for $\gamma$=10~(continuous line), 2~(dashed line), 1~(dotted-dashed line) and
0.1~(dotted line).}
 \label{fg:robin:momentum}

\item {Energy of a soliton colliding with the right (left) and left (right)
boundaries of a finite line subject to homogeneous Robin boundary conditions.
($\gamma$=10~(continuous line), 2~(dashed line), 1~(dotted-dashed line),
0.1~(dotted line)).}
 \label{fg:robin:energy}

\item {Maximum amplitude (left) and location of the maximum amplitude
(right) of a soliton colliding with  the right boundary of a finite line
subject to periodic~(continuous line), \D~(dashed line) and \N~(dotted-dashed
line)  boundary conditions.}
 \label{fg:location:PND}

\item {Maximum amplitude (top) and location of the maximum amplitude
(bottom) of a soliton colliding with the right (left) and left (right)
boundaries of a finite line subject to Robin boundary conditions. ($\gamma$
= 10~(continuous line), 2~(dashed line), 1~(dotted-dashed line), 0.1~(dotted
line)).}
 \label{fg:location:robin}

\item {Phase diagrams of the amplitude at right boundary~(top) and of the
soliton's maximum amplitude~(bottom) for the \N~(continuous line), \D~(dashed
line) and \P~(dotted-dashed line) problems. Note that only one collision with the
right boundary is presented, $L$=25 and $\Delta x$=0.0625.}
 \label{fg:phase:PND}

\item {Phase diagrams of the amplitude at the left (left) and right (right)
boundaries for the collision of a soliton with homogeneous \R\ \bc s.  Note
that only one collision with each boundary  is presented. ($L$=25, $\Delta
x$=0.0625, and $\gamma$ = 0.1~(continuous line), 1~(dashed line) and
10~(dotted-dashed line)).}
 \label{fg:phase:robin1}

\item {Phase diagrams of the amplitude at the left
(left) and right (right) boundaries of a finite line subject to homogeneous
\R\ boundary conditions. Note that five collisions with each
boundary are presented. ($\gamma$ = 1~(continuous
line) and  $-1$~(dashed line)).}
 \label{fg:phase:robin2}

\item {Phase diagram of the maximum
amplitude for  a \R\ problem. Note that two collisions, one with each
boundary, are presented, $L$=25 and $\Delta x$=0.0625. ($\gamma$ =
0.1~(continuous line), 1~(dashed line) and 10~(dotted-dashed line)).}
 \label{fg:lobes:robin1}

\item {Phase diagram of the maximum
amplitude for  a \R\ problem. Note that two collisions, one with each
boundary, are presented, $L$=25 and $\Delta x$=0.0625. ($\gamma$ =
1.125~(continuous line), 1.25~(dashed line) and 1.7~(dotted-dashed line)).}
 \label{fg:lobes:robin2}

\item {Areas of the phase diagrams corresponding to the amplitude at the
left (top left) and right (top right) boundaries and to the soliton's maximum
amplitude (bottom) as a function of the number of  collisions for the \R\
problem. Odd (even) collisions correspond to collisions of the soliton with the
right (left) boundary. (The symbols *, o, +, and x correspond to $\gamma$ = 10,
1.7, 1, and 0.6, respectively).}
 \label{fg:areas}

\item {Areas of the phase diagrams corresponding to the amplitude at the left
(top left) and right (top right)  boundaries and to the maximum amplitude of the
soliton for the sixth collision with the left (bottom left) and right (bottom
right) boundaries as functions of $\gamma$. The values of $\gamma$ = 0.1,
0.2, 0.4, 0.6, 0.8, 0.9, 1.0, 1.05, 1.075, 1.1, 1.125, 1.25, 1.7, 1.8, 1.9, 2.0,
2.5, 5.0 and 10 are represented by the natural numbers (gamma codes) from 1 to
19.}
 \label{fg:pitch}

\end{enumerate}

\newpage
\thispagestyle{empty}
\begin{textblock*}{\paperwidth}(0mm,0mm)\vspace{5cm}
   \noindent\includegraphics[width=\paperwidth]{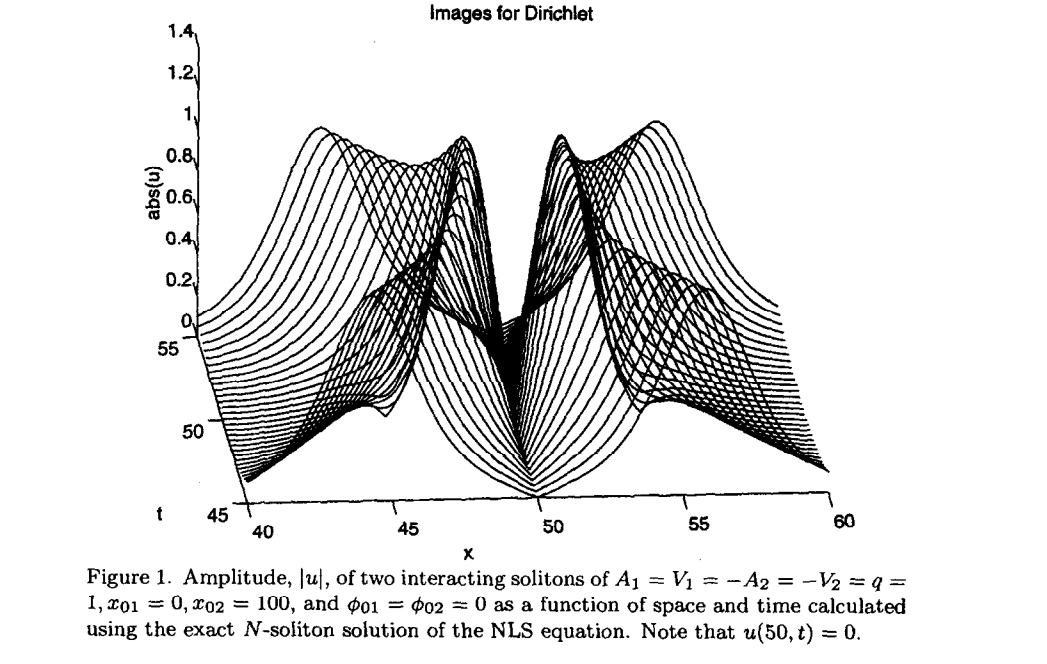}
\end{textblock*}
\mbox{}\newpage

\newpage
\thispagestyle{empty}
\begin{textblock*}{\paperwidth}(0mm,0mm)\vspace{5cm}
   \noindent\includegraphics[width=\paperwidth]{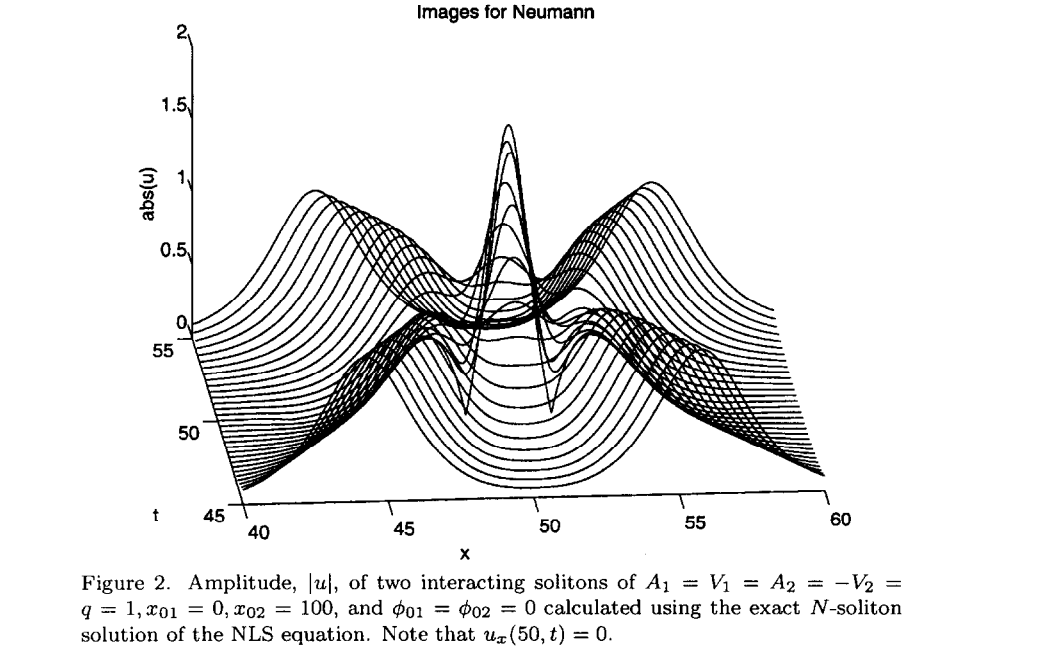}
\end{textblock*}
\mbox{}\newpage

\newpage
\thispagestyle{empty}
\begin{textblock*}{\paperwidth}(0mm,0mm)\vspace{5cm}
   \noindent\includegraphics[width=\paperwidth]{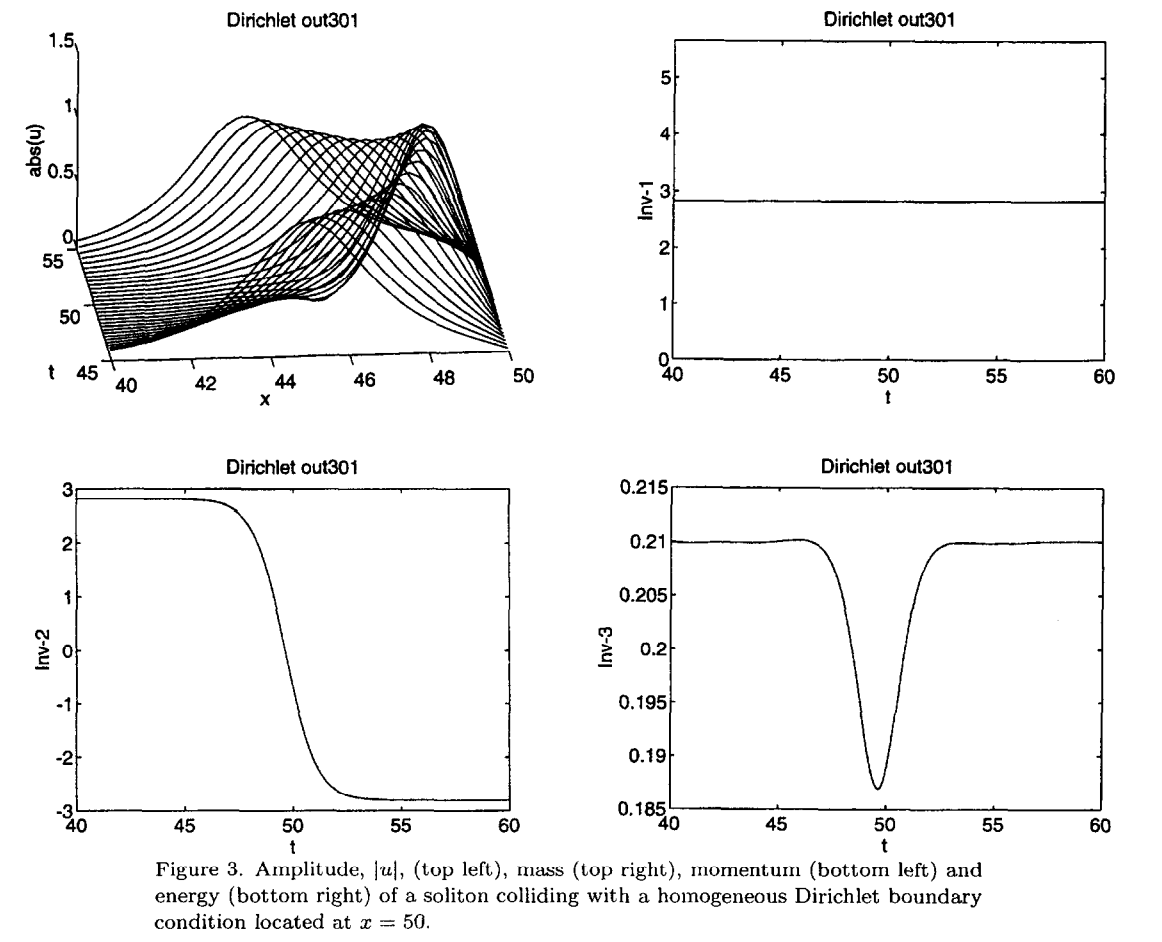}
\end{textblock*}
\mbox{}\newpage

\newpage
\thispagestyle{empty}
\begin{textblock*}{\paperwidth}(0mm,0mm)\vspace{5cm}
   \noindent\includegraphics[width=\paperwidth]{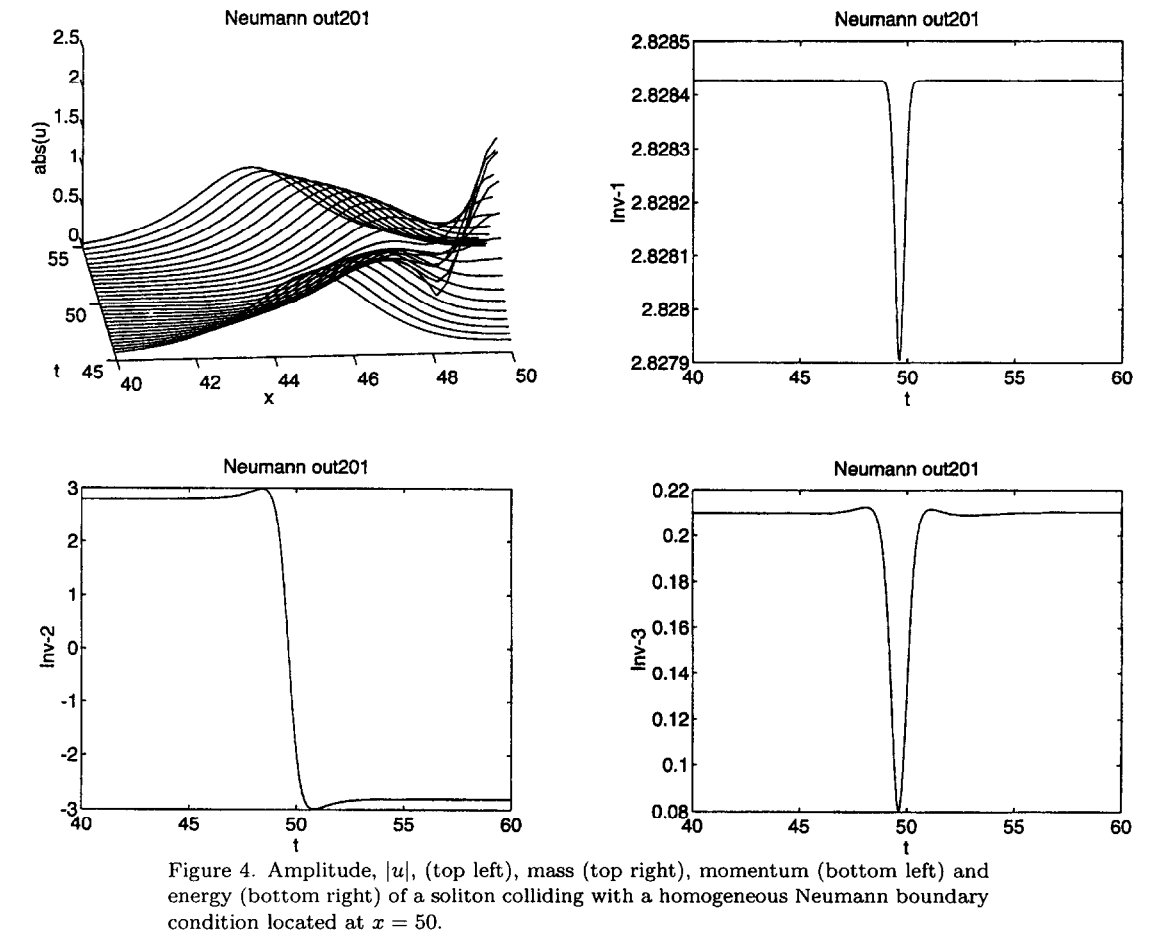}
\end{textblock*}
\mbox{}\newpage

\newpage
\thispagestyle{empty}
\begin{textblock*}{\paperwidth}(0mm,0mm)\vspace{5cm}
   \noindent\includegraphics[width=\paperwidth]{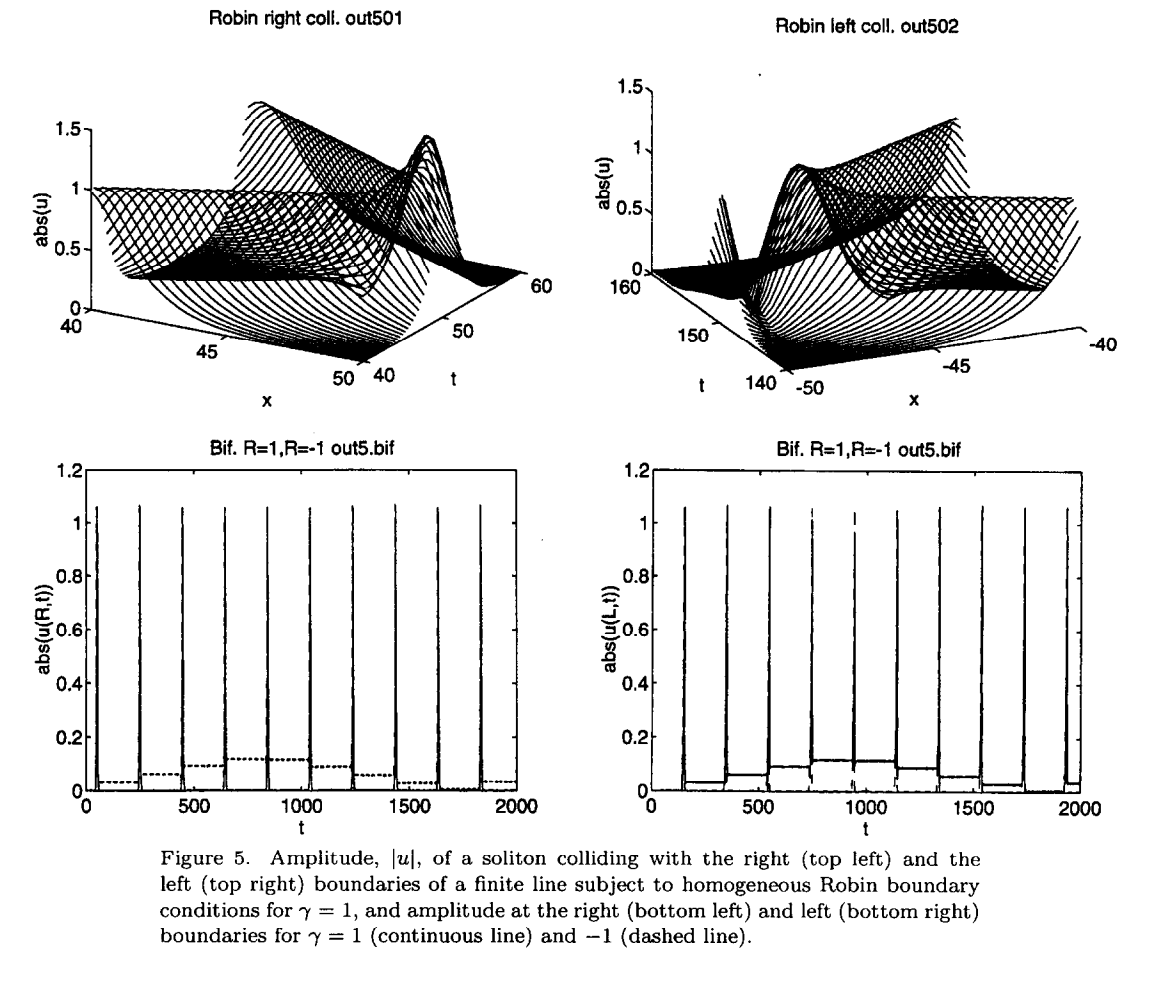}
\end{textblock*}
\mbox{}\newpage

\newpage
\thispagestyle{empty}
\begin{textblock*}{\paperwidth}(0mm,0mm)\vspace{5cm}
   \noindent\includegraphics[width=\paperwidth]{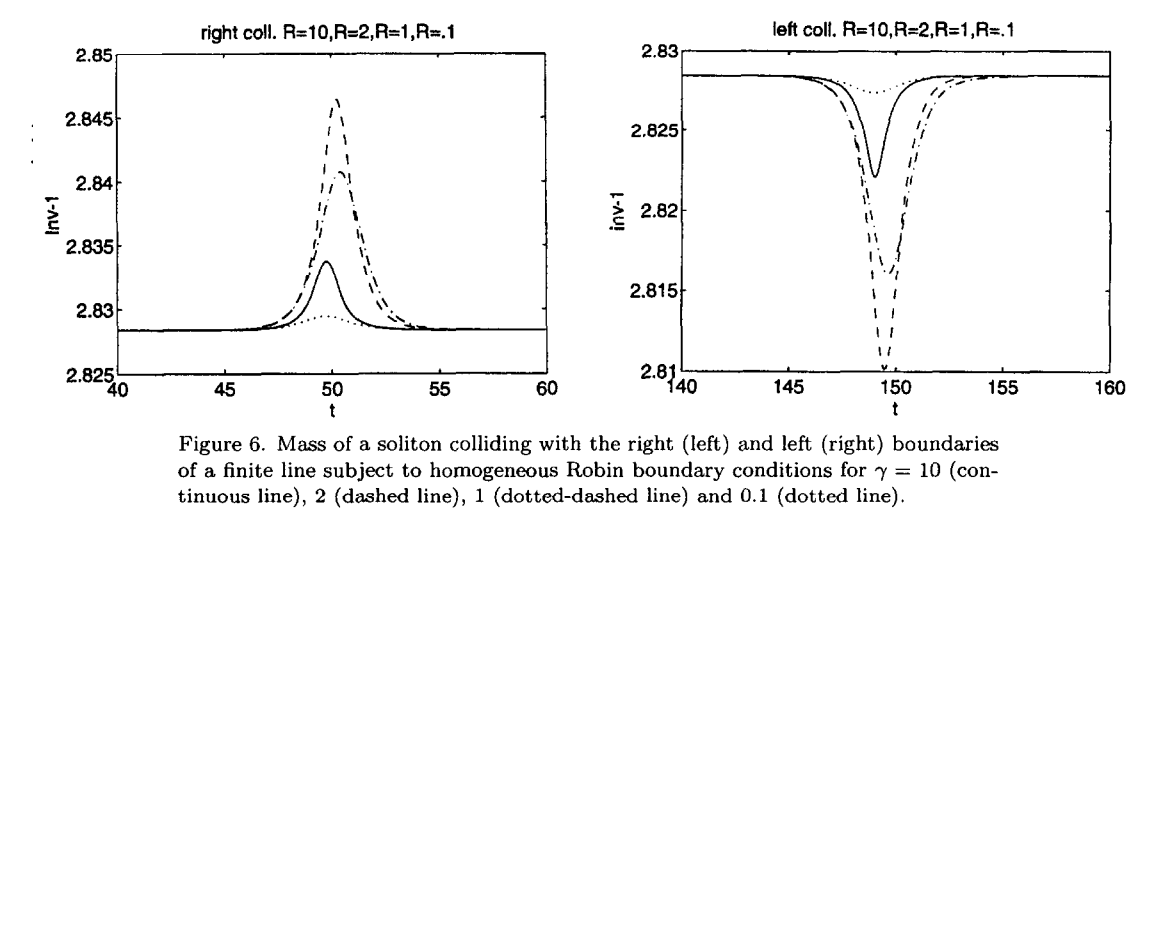}
\end{textblock*}
\mbox{}\newpage

\newpage
\thispagestyle{empty}
\begin{textblock*}{\paperwidth}(0mm,0mm)\vspace{5cm}
   \noindent\includegraphics[width=\paperwidth]{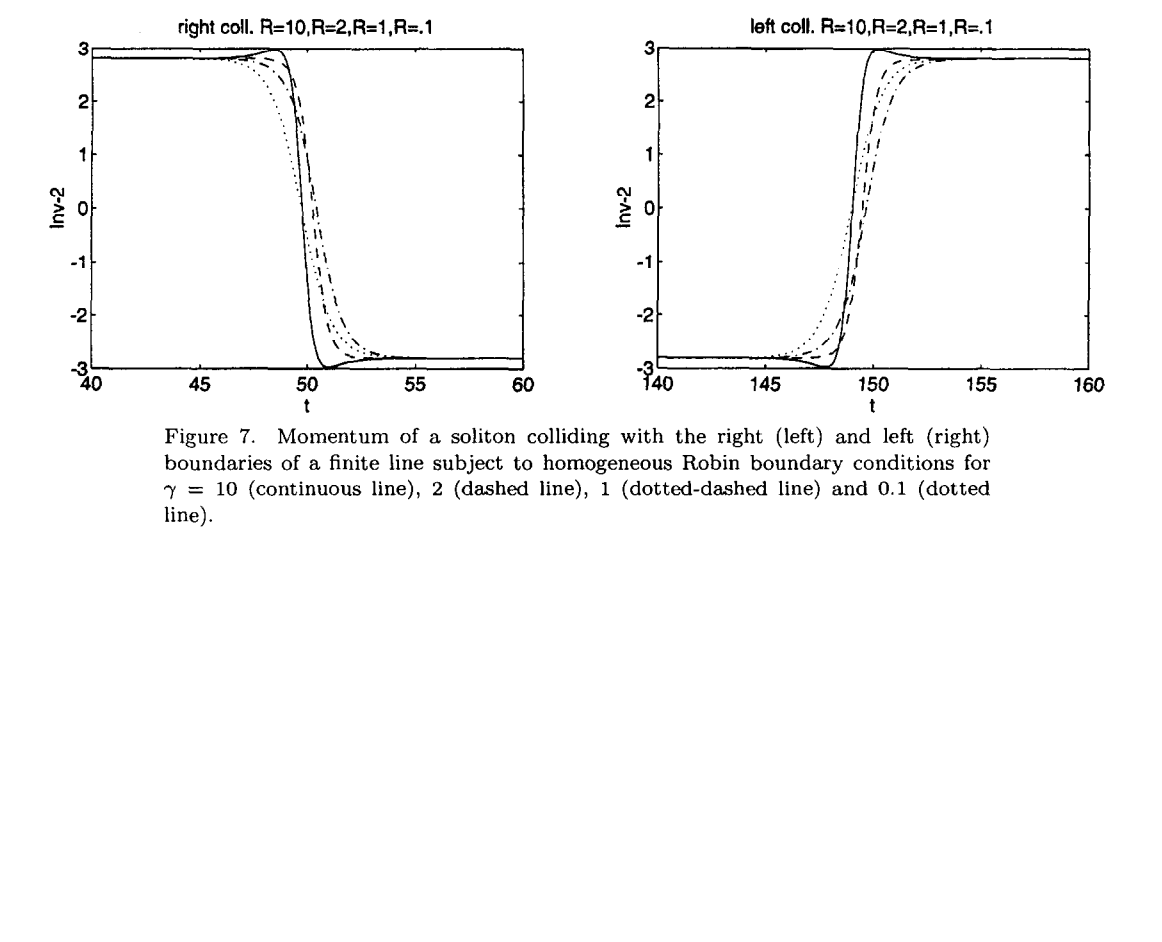}
\end{textblock*}
\mbox{}\newpage

\newpage
\thispagestyle{empty}
\begin{textblock*}{\paperwidth}(0mm,0mm)\vspace{5cm}
   \noindent\includegraphics[width=\paperwidth]{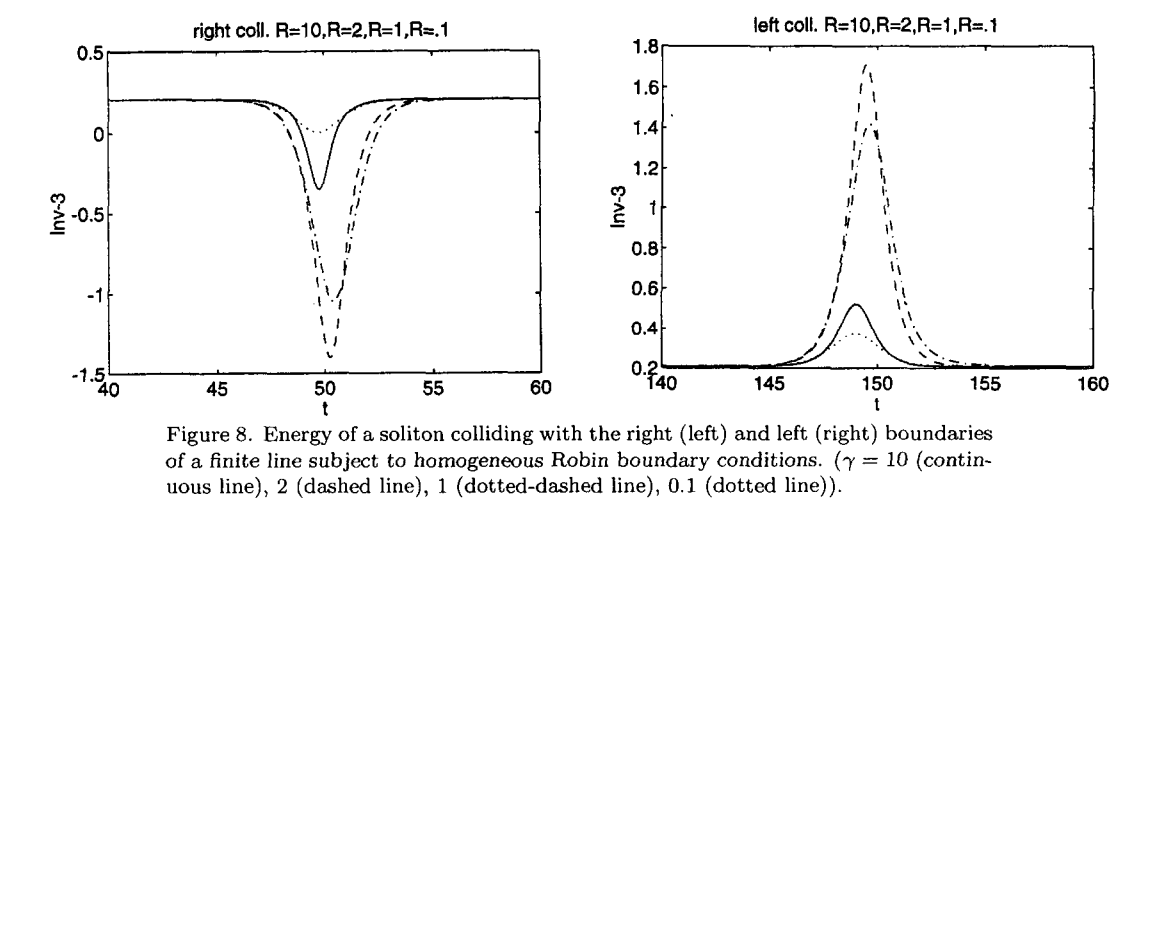}
\end{textblock*}
\mbox{}\newpage

\newpage
\thispagestyle{empty}
\begin{textblock*}{\paperwidth}(0mm,0mm)\vspace{5cm}
   \noindent\includegraphics[width=\paperwidth]{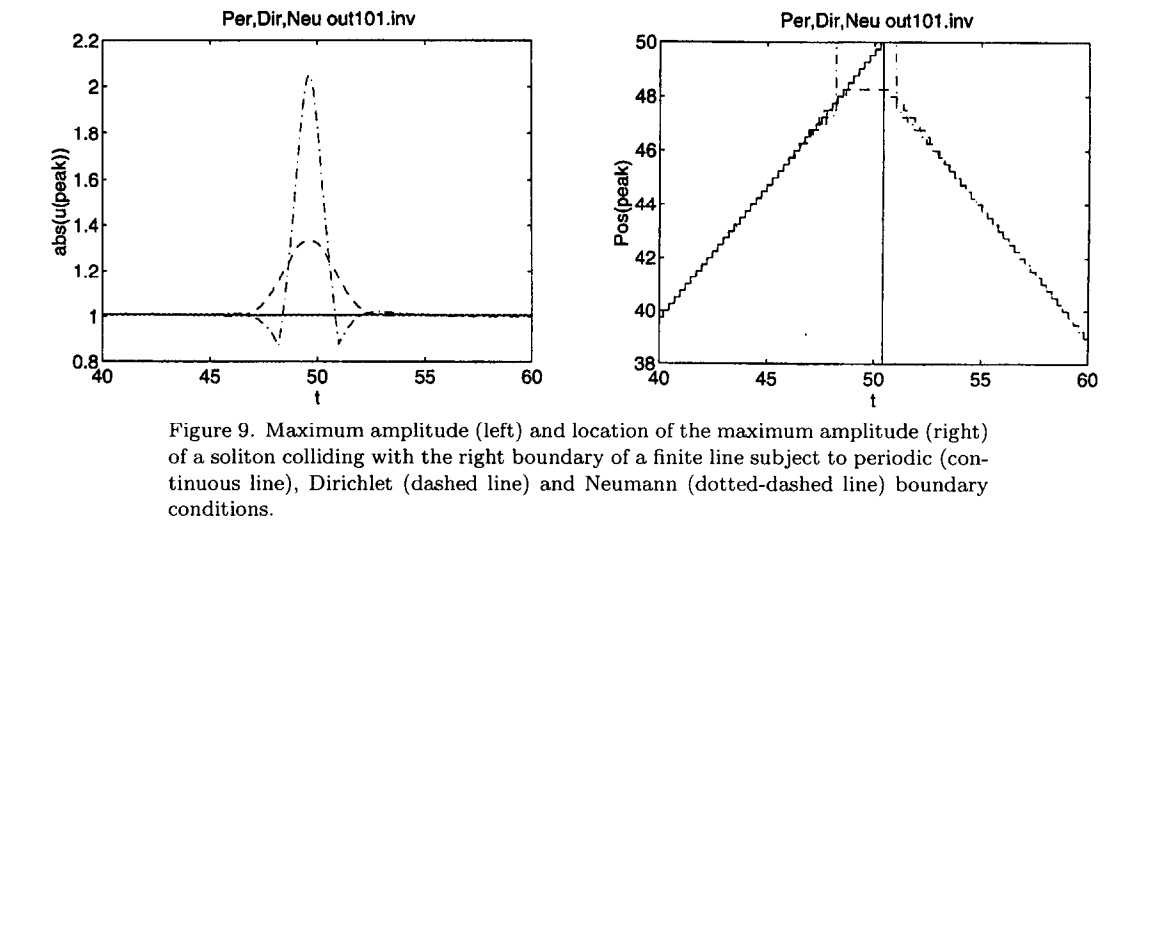}
\end{textblock*}
\mbox{}\newpage

\newpage
\thispagestyle{empty}
\begin{textblock*}{\paperwidth}(0mm,0mm)\vspace{5cm}
   \noindent\includegraphics[width=\paperwidth]{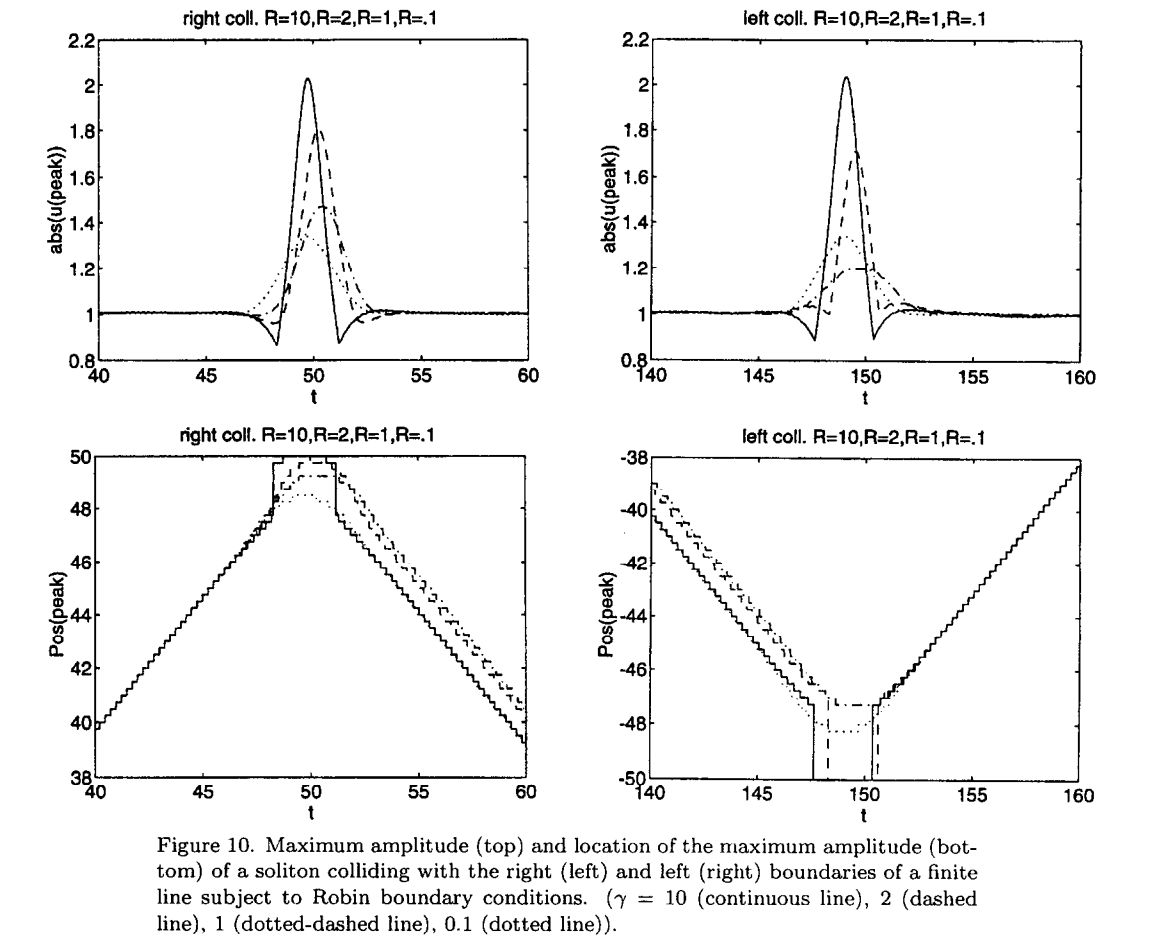}
\end{textblock*}
\mbox{}\newpage

\newpage
\thispagestyle{empty}
\begin{textblock*}{\paperwidth}(0mm,0mm)\vspace{5cm}
   \noindent\includegraphics[width=\paperwidth]{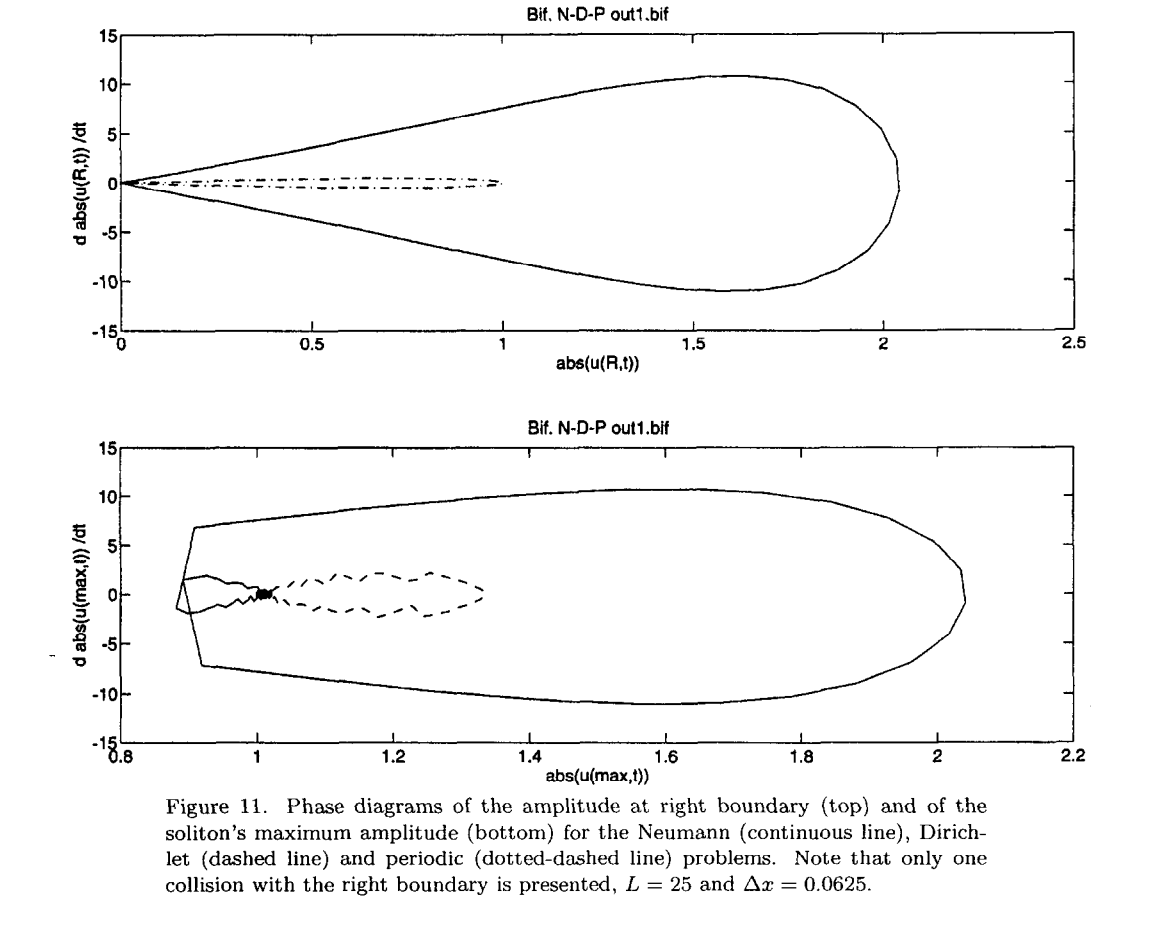}
\end{textblock*}
\mbox{}\newpage

\newpage
\thispagestyle{empty}
\begin{textblock*}{\paperwidth}(0mm,0mm)\vspace{5cm}
   \noindent\includegraphics[width=\paperwidth]{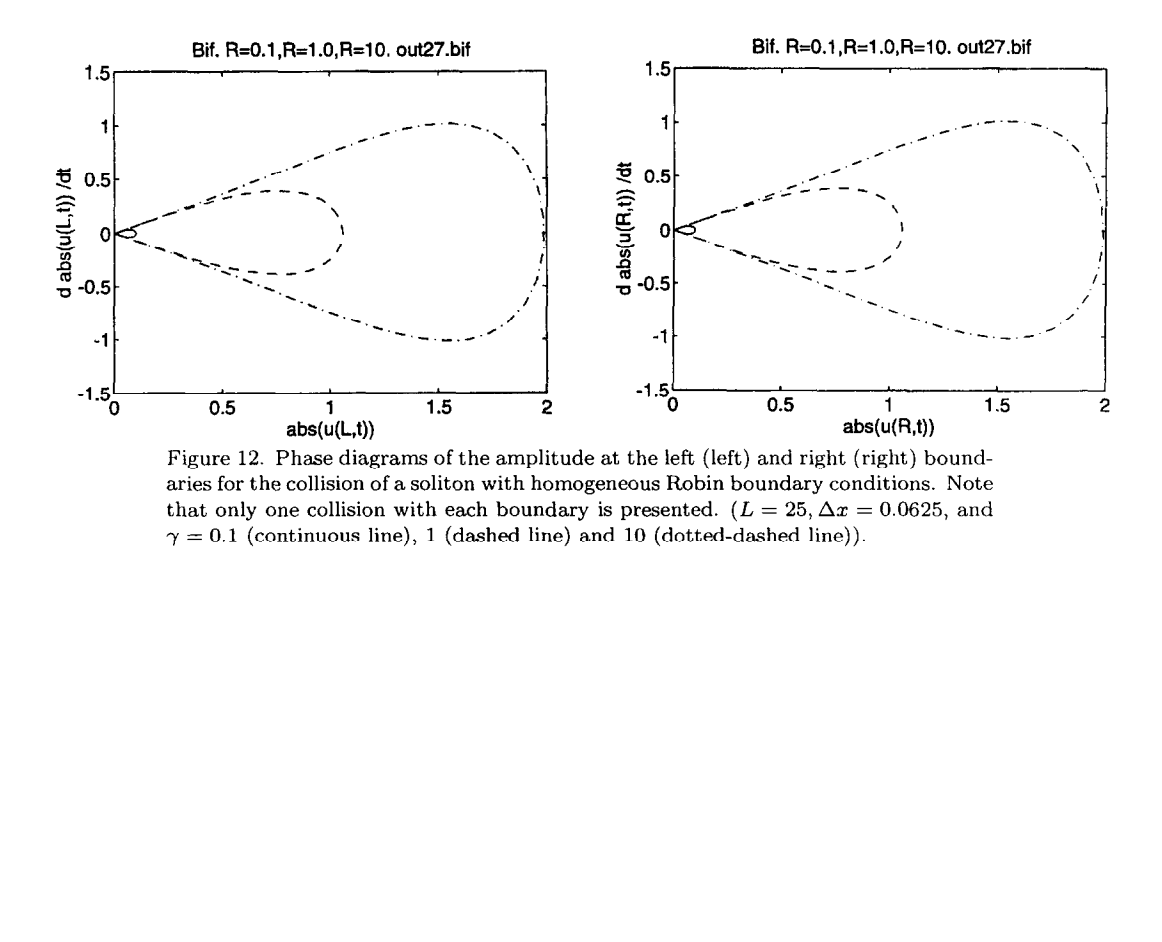}
\end{textblock*}
\mbox{}\newpage

\newpage
\thispagestyle{empty}
\begin{textblock*}{\paperwidth}(0mm,0mm)\vspace{5cm}
   \noindent\includegraphics[width=\paperwidth]{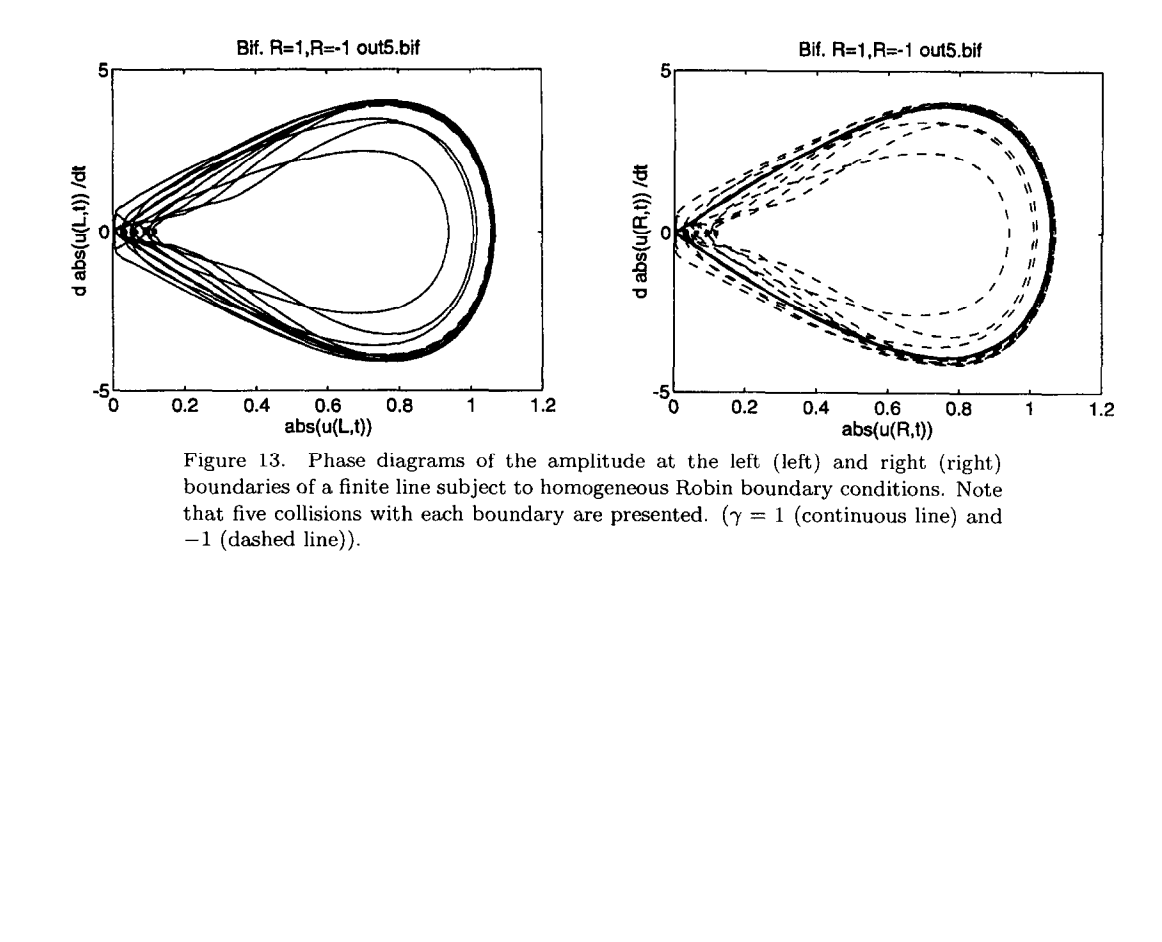}
\end{textblock*}
\mbox{}\newpage

\newpage
\thispagestyle{empty}
\begin{textblock*}{\paperwidth}(0mm,0mm)\vspace{5cm}
   \noindent\includegraphics[width=\paperwidth]{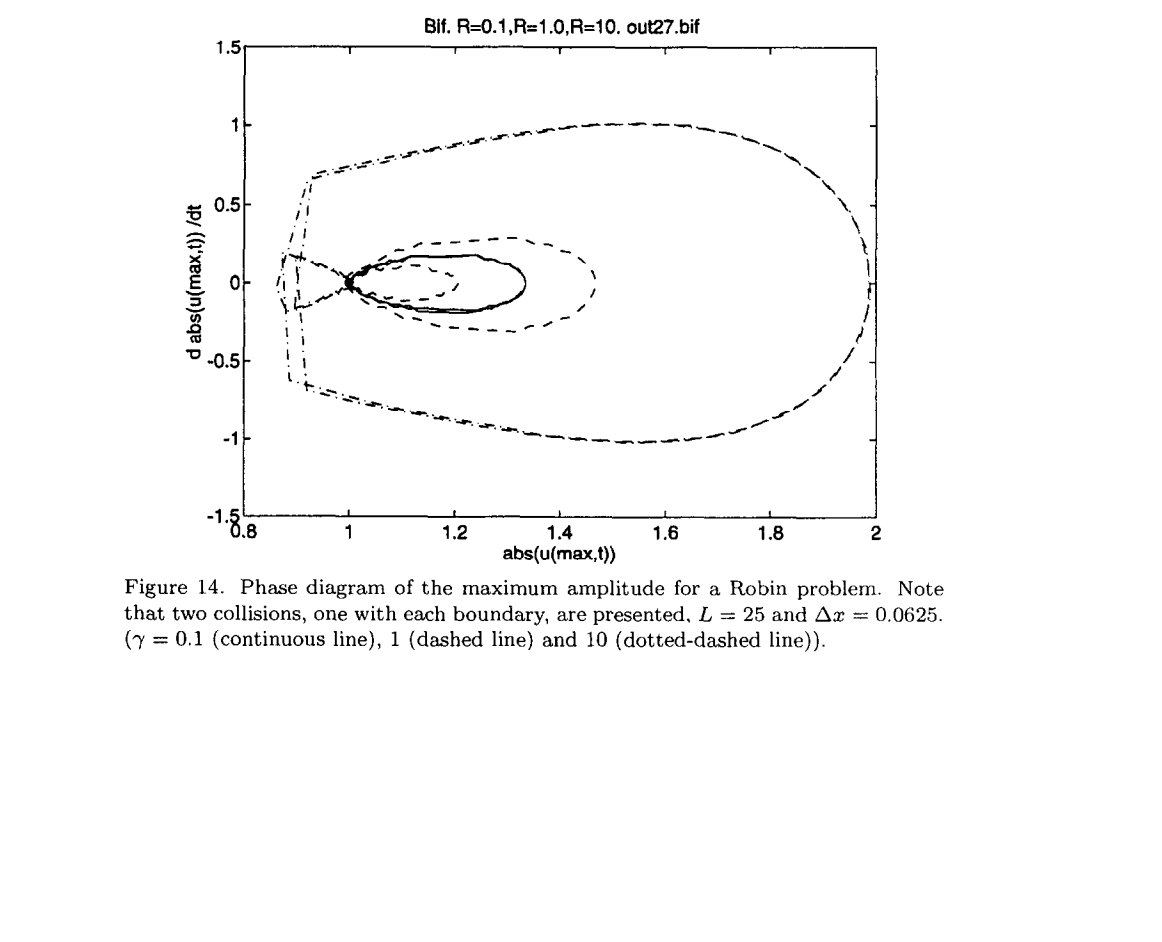}
\end{textblock*}
\mbox{}\newpage

\newpage
\thispagestyle{empty}
\begin{textblock*}{\paperwidth}(0mm,0mm)\vspace{5cm}
   \noindent\includegraphics[width=\paperwidth]{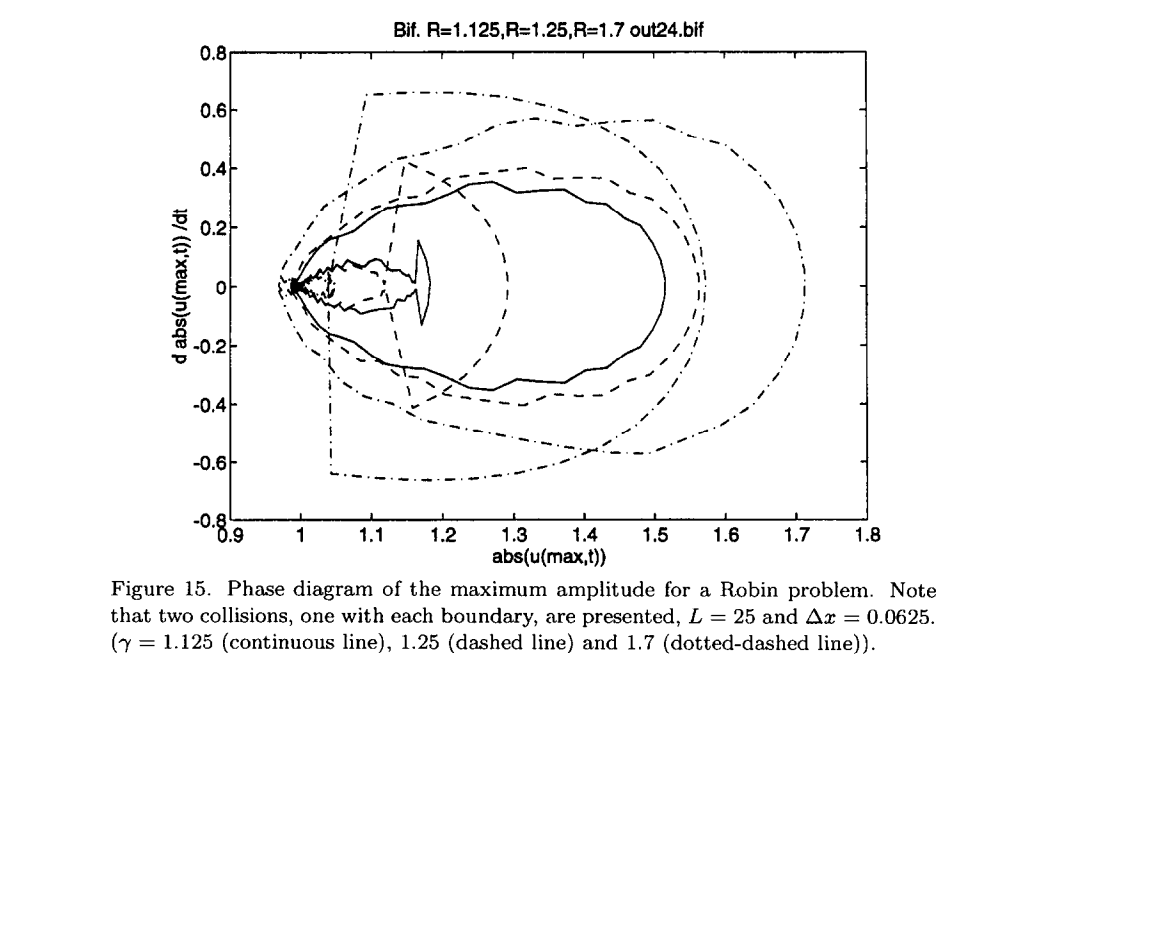}
\end{textblock*}
\mbox{}\newpage

\newpage
\thispagestyle{empty}
\begin{textblock*}{\paperwidth}(0mm,0mm)\vspace{5cm}
   \noindent\includegraphics[width=\paperwidth]{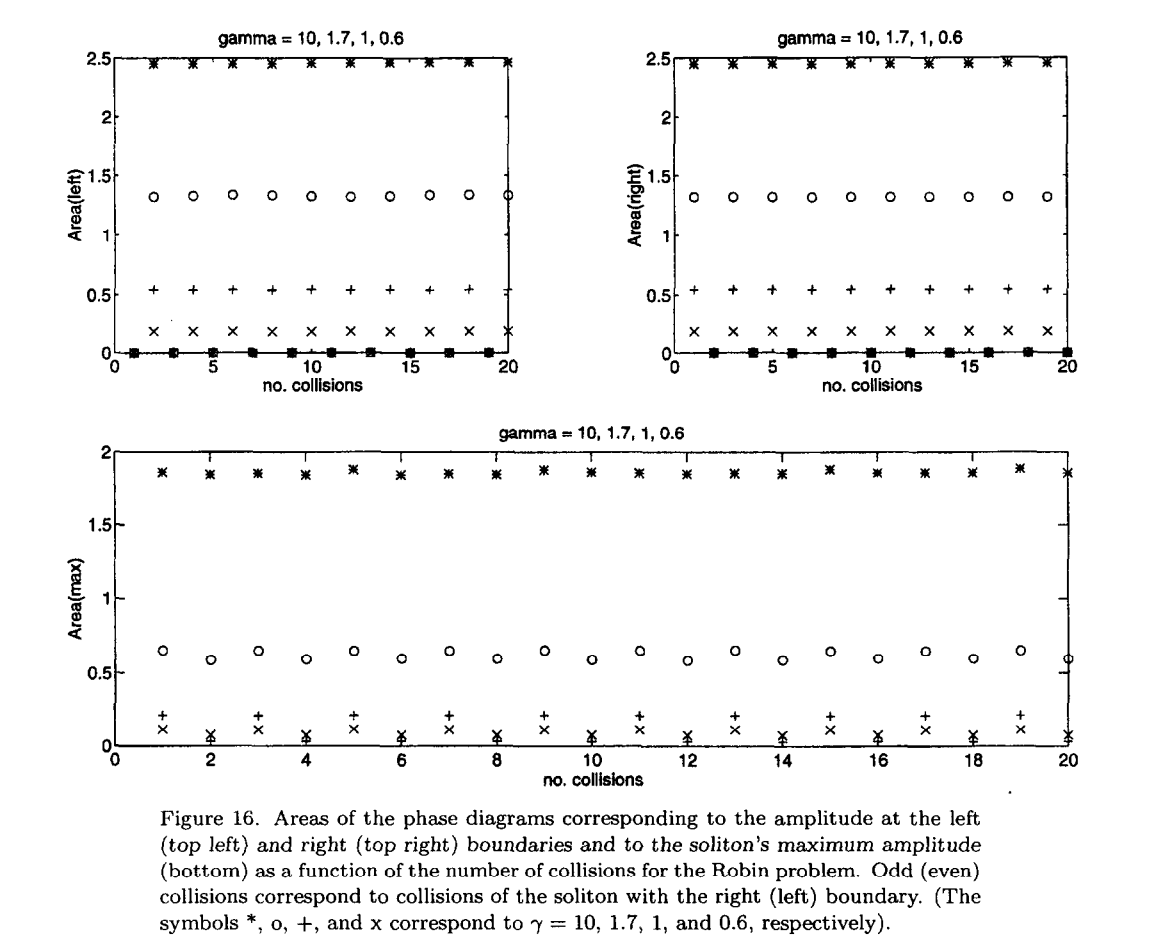}
\end{textblock*}
\mbox{}\newpage

\newpage
\thispagestyle{empty}
\begin{textblock*}{\paperwidth}(0mm,0mm)\vspace{5cm}
   \noindent\includegraphics[width=\paperwidth]{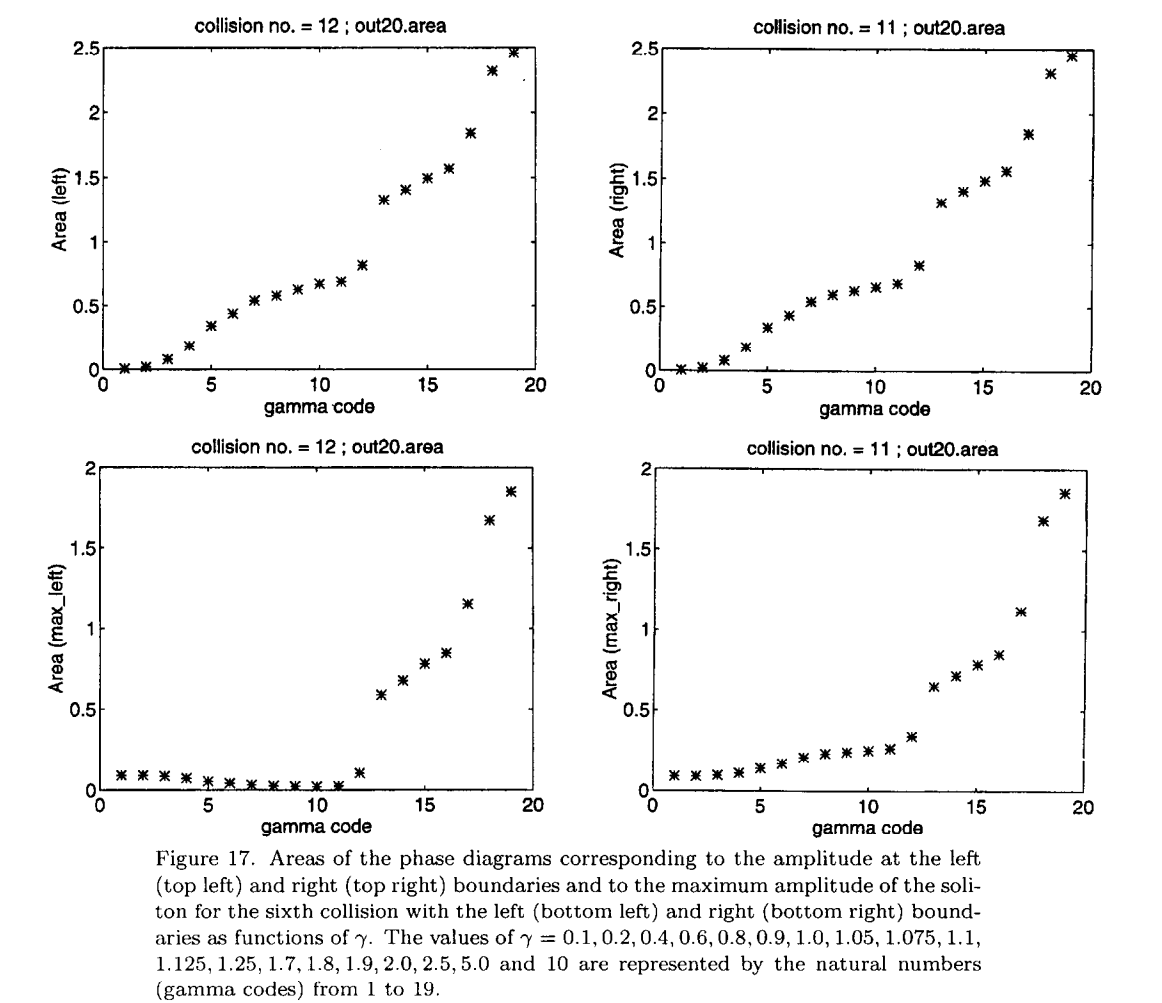}
\end{textblock*}
\mbox{}\newpage


\begin{thebibliography}{8}

\bibitem{Scot73} \paperrefiss {A. C. Scott, F. Y. F. Chu and W. McLaughlin} {The
Soliton---A New Concept in Applied Science} {Proc. IEEE} {61} {1443--1483}
{1973} {10}

\bibitem{Ablo81} \bookref {M. J. Ablowitz and H. Segur} {Solitons and the
Inverse Scattering Transform} {SIAM, Philadelphia} {4} {313--327} {1981}

\bibitem{Roge82} \bookref {C. Rogers and W. F. Shadwick} {B\"{a}cklund
Transformations and Their Applications} {Academic Press, New York} {1} {62--65}
{1982}

\bibitem{Calo91} \procref {F. Calogero} {Why Are Certain Nonlinear PDEs Both
Widely Applicable and Integrable?} {What Is Integrability?} {V. E. Zakharov}
{Springer-Verlag, Berlin} {1--61} {1991}

\bibitem{Zakh72} \paperrefiss {V. E. Zakharov and A. B. Shabat} {Exact Theory of
Two-Dimensional Self-Focusing and One-Dimensional Self-Modulation of Waves in
Nonlinear Media} {Sov. Phys. JETP} {34} {62--69} {1972} {1}

\bibitem{Weid86} \paperrefiss {J. A. C. Weideman and B. M. Herbst} {Split-Step
Methods for the Solution of the Nonlinear \schr\ Equation} {SIAM J. Numer.
Anal.} {23} {485--507} {1986} {3}

\bibitem{Sanz84} \paperref {J. M. Sanz-Serna} {Methods for the
Numerical Solution of the Nonlinear \schr\ Equation} {Math. Comp.} {43}
{21--27} {1984}

\bibitem{Taha84} \paperref {T. R. Taha and M. J. Ablowitz} {Classical and
Numerical Aspects of Certain Nonlinear Evolution Equations. II. Numerical,
Nonlinear \schr\ Equation} {J. Comput. Phys.} {55} {203--230} {1984}

\bibitem{Akri93} \paperref {G. D. Akrivis} {Finite Difference
Discretization of the Cubic \schr\ Equation} {IMA J. Num. Anal.} {13} {115--124}
{1993}

\bibitem{Grif84} \paperref {D. F. Griffiths, A. R. Mitchell and J. Ll. Morris}
{A Numerical Study of the Nonlinear \schr\ Equation} {Computer Meths. App. Mech. Engrg.}
{45} {177--215} {1984}

\bibitem{Argy87} \paperref {J. Argyris and M. Haase} {An Engineer's Guide to
Soliton Phenomena: Application of the Finite Element Method} {Computer Meths.
App. Mech. Engrg.} {61} {71--122} {1987}

\bibitem{Sham90} \paperrefiss {A. B. Shamardan} {The Numerical Treatment of the
Nonlinear \schr\ Equation} {Computers Math. Applic.} {19} {67--73} {1990} {4}

\bibitem{Kaup85} \procref {D. J. Kaup} {Approximations for the Inverse
Scattering Transform} {Dynamical Problems in Soliton Systems} {S. Takeno}
{Springer-Verlag, Berlin} {12--22} {1985}

\bibitem{Ablo91a} \bookref {M. J. Ablowitz and P. A. Clarkson} {Solitons,
Nonlinear Evolution Equations and Inverse Scattering} {Cambridge University
Press, Cambridge} {8} {426--430} {1991}

\bibitem{Chu90} \procref {C. K. Chu and R. L. Chou} {Solitons Induced by
Boundary Conditions} {Advances in Applied Mechanics, Vol. 27\,} {J. W. Hutchinson
and T. Y. Wu} {Academic Press, Boston} {283--302} {1990}

\bibitem{Chu83} \paperref {C. K. Chu, L. W. Xiang and Y. Baransky}
{Solitary Waves Induced by Boundary Motion} {Comm. Pure Appl. Math.}
{36} {495--504} {1983}

\bibitem{Chou90} \paperrefiss {R. L. Chou and C. K. Chu}
{Solitons Induced by Boundary Conditions from the Boussinesq Equation} {Phys.
Fluids A} {2} {1574--1584} {1990} {9}

\bibitem{Kaup86} \paperref {D. J. Kaup and P. J. Hansen} {The Forced Nonlinear
\schr\ Equation} {Physica D} {18} {77--84} {1986}

\bibitem{Ablo75} \paperrefiss {M. J. Ablowitz and H. Segur} {The Inverse Scattering
Transform: the Semi-Infinite Interval} {J. Math. Phys.} {16} {1054--1056}
{1975} {5}

\bibitem{Foka89a} \paperref {A. S. Fokas} {An Initial-Boundary Value Problem for
the  Nonlinear \schr\ Equation} {Physica D} {35} {167--185} {1989}

\bibitem{Bikb91} \paperref {R. F. Bikbaev and V. O. Tarasov} {Initial Boundary
Value Problem for the Nonlinear \schr\ Equation} {J. Phys. A: Math. Gen.} {24}
{2507--2516} {1991}

\bibitem{Foka89b} \paperref {A. S. Fokas and M. J. Ablowitz} {Forced Nonlinear
Evolution Equations and the Inverse Scattering Transform} {Stud. Appl. Math.} {80}
{253--272} {1989}

\bibitem{Ma81} \paperref {Y.-C. Ma and M. J. Ablowitz} {The Periodic Cubic
Schr\"{o}dinger Equation} {Stud. Appl. Math.} {65} {113--158} {1981}

\bibitem{Osbo93} \paperref {A. R. Osborne} {The Hyperelliptic Inverse Scattering
Transform for the Periodic, Defocusing Nonlinear Schr\"{o}dinger Equation} {J.
Comput. Phys.} {109} {93--107} {1993}

\bibitem{Boyd90} \procref {J. P. Boyd} {New Directions in Solitons and Nonlinear
Periodic Waves: Polycnoidal Waves, Imbricated Solitons, Weakly Nonlocal Solitary
Waves, and Numerical Boundary Value Algorithms} {Advances in Applied Mechanics, Vol.
27} {J. W. Hutchinson and T. Y. Wu} {Academic Press, Boston} {1--82} {1990}

\bibitem{Chri85} \procref {P. L. Christiansen} {Solitons and Chaos in the
Sine-Gordon  System} {Dynamical Problems in Soliton Systems} {S. Takeno}
{Springer-Verlag, Berlin} {258--261} {1985}

\bibitem{Siro90} \paperref {L. Sirovich, J. D. Rodriguez and B. Knight} {Two
Boundary Value Problems for the Ginzburg-Landau Equation} {Physica D} {43}
{63--76} {1990}

\bibitem{Miri82} \paperref {R. M. Mirie and C. H. Su} {Collisions Between Two
Solitary Waves. Part 2. A Numerical Study} {J. Fluid Mech.} {115} {475--492} {1982}

\bibitem{Ablo91b} \paperref {M. J. Ablowitz and B. M. Herbst and J. A. C.
Weideman} {Dynamics of Semi-Discretizations of the Defocusing Nonlinear
{Schr\"{o}dinger} Equation} {IMA J. Numer. Anal.} {11} {539--552} {1991}

\bibitem{Gord83} \paperrefiss {J. P. Gordon} {Interaction Forces Among Solitons in
Optical  Fibers} {Optics Lett.} {8} {596--598} {1983} {11}

\bibitem{Ben84} \paperrefiss {G. Ben-Yu} {The Convergence of a Numerical Method for
the Nonlinear Schr\"{o}dinger Equation} {J. Comput. Math.} {4} {121--130}
{1984} {2}

\bibitem{Rals78} \bookref {A. Ralston and Ph. Rabinowitz} {A First
Course in Numerical  Analysis} {2nd ed., McGraw-Hill Book Co., New York} {?}
{118--130} {1978}

\bibitem{Thom86} \bookref {J. M. T. Thompson and H. B. Stewart} {Nonlinear
Dynamics and Chaos} {John Wiley and Sons, Chichester} {4} {84--87} {1986}





\end{thebibliography}
\end{document}